\documentclass[twocolumn,prc,amsmath,amssymb,aps,showpacs,superscriptaddress,floatfix]{revtex4-1}
\usepackage{graphicx}
\usepackage{epstopdf}
\usepackage{dcolumn} % Align table columns on decimal piont
\usepackage{bm} % Bold Math
\usepackage{multirow}
\usepackage{longtable}
\usepackage{tensor}
\usepackage{threeparttable}

\newcommand{\nuc}[2]{\ifmmode {{}^{#1}\text{#2}} \else
  {${}^{#1}\text{#2}$}\fi}

\newcommand{\Beleven}{\nuc{11}{B}}
\newcommand{\Ctwelve}{\nuc{12}{C}}
\newcommand{\Celeven}{\nuc{11}{C}}

\newcommand{\lamb}[2]{\ifmmode {\tensor*[^{#1}_{\Lambda}]{\text{#2}}{}} \else
  {$\tensor*[^{#1}_{\Lambda}]{\text{#2}}{}$}\fi}

\newcommand{\BLtwelve}{\lamb{12}{B}}

\newcommand{\eepK}{\ifmmode (e,e^{\prime}K^+)\else $(e,e^{\prime}K^+)$\fi}
\newcommand{\Kpi}{\ifmmode (K^-,\pi^-)\else $(K^-,\pi^-)$\fi}
\newcommand{\piK}{\ifmmode (\pi^+,K^+)\else $(\pi^+,K^+)$\fi}
\newcommand{\gK}{\ifmmode (\gamma,K^+)\else $(\gamma,K^+)$\fi}
\newcommand{\Kpig}{\ifmmode (K^-,\pi^-\gamma)\else $(K^-,\pi^-\gamma)$\fi}
\newcommand{\piKg}{\ifmmode (\pi^+,K^+\gamma)\else $(\pi^+,K^+\gamma)$\fi}

\newcommand{\etal}{\textit{et al.}}

\newcommand{\CHtwo}{\ifmmode {\text{CH}_2} \else
  {$\text{CH}_2$}\fi}

\newcommand{\vect}[1]{\boldsymbol{\mathbf{#1}}}

\newcommand{\unsim}{\mathord{\sim}}

\begin{document}

\title{The experiments with the High Resolution Kaon Spectrometer at JLab
  Hall C and the new spectroscopy of \lamb{12}{B} hypernuclei}

\newcommand*{\HAMPTON}{Department of Physics, Hampton University, Virginia 23668, USA}
\newcommand*{\JLAB}{Thomas Jefferson National Accelerator Facility, Newport News, Virginia 23606, USA}
\newcommand*{\TOHOKU}{Graduate School of Science, Tohoku University, Sendai, Miyagi 980-8578, Japan}
\newcommand*{\ZAGREB}{Department of Physics \& Department of Applied Physics, University of Zagreb, HR-10000 Zagreb, Croatia}
\newcommand*{\HOU}{Department of Physics, University of Houston, Houston, Texas 77204, USA}
\newcommand*{\SUAGM}{Escuela de Ciencias y Tecnologia, Universidad Metropolitana, San Juan, Puerto Rico}
\newcommand*{\FIU}{Department of Physics, Florida International University, Miami, Florida 33199, USA}
\newcommand*{\MAINZ}{Institute f\"{u}r Kernphysik, Johannes Gutenberg-Universit\"{a}t Mainz, D-55099 Mainz, Germany}
\newcommand*{\NCSU}{Department of Physics, North Carolina A\&T State University, Greesboro, North Carolina 27411, USA}
\newcommand*{\YERPHY}{Yerevan Physics Institute, Yerevan 0036, Armenia}
\newcommand*{\MARYLAND}{Department of Physics, University of Maryland, College Park, Maryland 20742, USA}
\newcommand*{\CNU}{Department of Physics, Christopher Newport University, Newport News, VA 23606, USA}
\newcommand*{\UVA}{University of Virginia, Charlotesville, Virginia 22904, USA}
\newcommand*{\BARI}{Istituto Nazionale di Fisica Nucleare, 
Sezione di Bari and University of Bari, I-70126 Bari, Italy}
\newcommand*{\SUNO}{Department of Physics, Southern University at New Orleans, New Orleans, LA 70126, USA}
\newcommand*{\NCW}{Department of Physics, University of North Carolina Wilmington, Wilmington, NC 28403, USA}
\newcommand*{\INFN}{INFN, Sezione Sanit\`{a} and Istituto Superiore di Sanit\`{a}, 00161 Rome, Italy}
\newcommand*{\CSU}{Physics and Astronomy Department, California State
University, Sacramento CA 95819, USA}
\newcommand*{\RIKEN}{Institute for Physical and Chemical Research
(RIKEN), Wako, Saitama 351-0198, Japan}
\newcommand*{\LANZHOU}{Nuclear Physics Institute, Lanzhou University, Lanzhou, Gansu, 730000, China}
\newcommand*{\LATECH}{Department of Physics, Louisiana Tech
University, Ruston, Louisiana 71272, USA}
\newcommand*{\YAMAGATA}{Faculty of Science, Yamagata University, Yamagata 990-8560, Japan}
\newcommand*{\BNL}{Brookhaven National Laboratory, Upton, NY 11973, USA}
\newcommand*{\MISSSTATE}{Mississippi State University, Mississippi State, Mississippi 39762, USA}
\newcommand*{\JMU}{Department of Physics, James Madison University, Harrisonburg, Virginia 22807, USA}
\newcommand*{\LEXINGTON}{Department of Physics \& Astronomy, Virginia Military Institute, 
Lexington, VA 24450, USA}
\newcommand*{\KEK}{Institute of Particle and Nuclear Studies, KEK, Tsukuba, Ibaraki 305-0801, Japan}
\newcommand*{\XAVIER}{Department of Physics, Xavier University of Louisiana, New Orleans, LA, USA}
\newcommand*{\OSAKA}{Laboratory of Physics, Osaka Electro-Communication 
University, Neyagawa, Osaka 572-8530, Japan}
\newcommand*{\YUKAWA}{Yukawa Institute for Theoretical Physics, Kyoto University, Kyoto 606-8502, Japan}

\author{L.~Tang}
\email{Corresponding author. tangl@jlab.org}
\affiliation{\HAMPTON}
\affiliation{\JLAB}
\author{C.~Chen}
\affiliation{\HAMPTON}
\author{T.~Gogami}
\affiliation{\TOHOKU}
\author{D.~Kawama}
\affiliation{\TOHOKU}
\author{Y.~Han}
\affiliation{\HAMPTON}
\author{L.~Yuan}
\affiliation{\HAMPTON}
\author{A.~Matsumura}
\affiliation{\TOHOKU}
\author{Y.~Okayasu}
\affiliation{\TOHOKU}
\author{T.~Seva}
\affiliation{\ZAGREB}
\author{V.~M.~Rodriguez}
\affiliation{\HOU}
\affiliation{\SUAGM}
% ...

\collaboration{HKS (JLab E05-115 and E01-011) Collaborations}
\noaffiliation

\begin{abstract}
Since the pioneering experiment, E89-009 studying hypernuclear 
spectroscopy using the {\eepK} reaction was completed, two additional 
experiments, E01-011 and E05-115, were performed at Jefferson Lab. 
These later experiments used a modified experimental design, the 
``tilt method'', to dramatically suppress the large electromagnetic 
background, and allowed for a substantial increase in luminosity.  
Additionally, a new kaon spectrometer, HKS (E01-011), a new electron 
spectrometer, HES and a new splitting magnet (E05-115), were added to 
produce new data sets of precision, high-resolution hypernuclear 
spectroscopy. All three experiments obtained a spectrum for \lamb{12}{B},
which is the most characteristic $p$-shell hypernucleus and is commonly 
used for calibration.  Independent analyses of these different experiments 
demonstrate excellent consistency and provide the clearest level 
structure to date of this hypernucleus as produced by the {\eepK} 
reaction. This paper presents details of these experiments, and the 
extraction and analysis of the observed \lamb{12}{B} spectrum.
\end{abstract}

\date{{\today}}

\maketitle

\section{\label{sec:intro}Introduction}

 Spectroscopic investigation of $\Lambda$ hypernuclei is a unique 
method which provides invaluable information on many-body baryonic 
systems by inserting a new degree of freedom, ``strangeness'', into 
the nucleus.  Since the $\Lambda$ is not Pauli-blocked, it can occupy 
any single-particle state, providing a distinguishable probe of the 
nuclear interior~\cite{Povh87,Chrien89,Bando90}. Therefore, new 
nuclear structures or unknown properties of the baryonic interaction, 
which cannot be seen from the investigation of ordinary nuclei with
conventional probes, may manifest themselves in hypernuclei, providing
indispensable information on flavor SU(3) for baryonic matter. 
In addition, a study of hypernuclear spectra provides the only 
practical way to study the $\Lambda N$ interaction, as 
$\Lambda N$ scattering experiments are technically difficult or 
impossible.  

 Aside from strangeness, another important feature is the absence of 
isospin (I = 0) of the $\Lambda$.  As isospin conservation
prevents one-pion-exchange (OPE) 
in the $\Lambda N$ interaction, the 
long range OPE component is absent and thus the $\Lambda N$ interaction is 
more sensitive to short range components of the strong interaction
than the nucleon-nucleon interaction.
Since $\Lambda$ decays weakly and has a relatively long lifetime 
($\unsim 260$ ps), the spectroscopy of $\Lambda$ hypernuclei features narrow 
states commonly described by coupling low-lying nuclear-hole states to
$\Lambda$ single particle states with widths ranging from a few to 
$\unsim 100$ keV.
This makes detailed spectroscopic 
studies possible.

 A phenomenological approach to $p$-shell $\Lambda$ hypernuclei 
introduces a two-body effective potential~\cite{Dalitz78,Millener85} 
in form of
\begin{equation}
  \label{eqn:eq1}
  \begin{split}
    V_{\Lambda N} = V_0 (r) &+ V_{\sigma}(r) \vect{s}_{\Lambda} \cdot
    \vect{s}_N + V_{\Lambda}(r) \vect{l}_{\Lambda N} \cdot
    \vect{s}_{\Lambda}\\ &+ V_{\Lambda}(r) \vect{l}_{\Lambda N} \cdot
    \vect{s}_N + V_T (r) S_{12}\; ,
  \end{split}
\end{equation}
where $S_{12} = 3(\vect{\sigma}_{\Lambda} \cdot \vect{r}/r)
(\vect{\sigma}_N \cdot \vect{r}/r) - \vect{\sigma}_{\Lambda} \cdot
\vect{\sigma}_N$.  Low-lying levels of $p$-shell hypernuclei can be 
described with radial integrals over the $s_{\Lambda}$$p_N$ wave 
function for each of the five terms in Eq.~(\ref{eqn:eq1}).  A set of 
these integrals, denoted as $\bar{V}$, $\Delta$, $S_{\Lambda}$, $S_N$ 
and $T$, can be determined from selected $p$-shell $\Lambda$-hypernuclear 
spectroscopy and then used to fit the $\Lambda N$ interactions.  

 The other approach applies a G-matrix derived from models 
(Nijmegen~\cite{Rijken99,Rijken06,Rijken10,Yamamoto13} or 
J\"{u}lich~\cite{Haidenbauer05,Haidenbauer13}) which describe the
Baryon-Baryon interactions including 
the free $\Lambda N$ interactions. When using this more direct 
description, the properties of $\Lambda N$ interaction models can be 
explored. However, high precision spectroscopy is required with either 
approach in order to obtain reliable information on the unique 
characteristics of the $\Lambda N$ interaction.

 Traditionally the spectroscopy of $\Lambda$ hypernuclei were obtained 
using beams of pions or kaons, either stopped in a target or in-flight.  
In reactions such as $(K^-, \pi^-)$ or $(\pi^+, K^+)$,  a nucleon in a 
target was replaced by a $\Lambda$.  However, the resolution using mesonic 
beams was limited to about or more than 1.5 MeV (FWHM) due to the fact 
that these beams are produced by reactions of a primary beam on a 
production target and thus are limited in intensity.  To compensate for the
low beam intensity, thick targets ($>500~\text{mg}/\text{cm}^2$)
were used broadening the resolution by the uncertainty in energy
loss.   Weakly excited states, particularly low-lying 
states, were difficult to resolve and their binding energies inaccurately 
extracted.  Yet in many cases the weakly produced states are quite
important when comparing an experiment to theoretical calculations. 
For example, the recent high precision $\gamma$-transition spectroscopy 
experiments at KEK and AGS reported a total of 22 precisely measured 
level transitions for several $p$-shell hypernuclei~\cite{Hashimoto06}.
These results enabled a detailed theoretical study of $p$-shell 
$\Lambda$ hypernuclei. New values of the integrals given in 
Eq.~(\ref{eqn:eq1}) were extracted, as well as contributions from each 
term to the binding energies~\cite{Millener12}.  However, gamma 
transition energies cannot provide information on ground state binding 
energies. 

 Electroproduction using the {\eepK} reaction with intense beams at the 
Jefferson Lab accelerator provides a unique opportunity to study high 
precision hypernuclear spectroscopy.  An energy resolution of $\unsim 500$ 
keV (FWHM) can be achieved using a combination of (1) the small 
emittance of the electron beam, (2) the excellent momentum resolution 
using precision spectrometers for the scattered electron and produced
kaon, (3) the precision measurement of the scattering angles, and (4) 
the thin target foils minimizing target straggling and radiative corrections. 
On the other hand, the experimental design must accommodate high luminosity, 
potential backgrounds, and precise calibration of the spectrometers. 

  Electroproduction brings in additional new features to the overall 
investigation of hypernuclei.  The {\eepK} reaction produces a $\Lambda$ 
from a proton in the nucleus creating a proton hole in the core to which 
the $\Lambda$ couples.  This can produce either mirror hypernuclei to 
those produced by the hadronic reactions {\Kpi} and {\piK}, or states 
with different isospin. Thus, electroproduction produces neutron-rich
hypernuclei that are suitable candidates to investigate $\Lambda N-\Sigma N$ 
coupling and the effective $\Lambda NN$ three-body force.  Furthermore, 
electroproduction involves large spin-flip transition amplitudes from 
the initial nuclear to the final hypernuclear states. Still, the 
non-spin-flip amplitude remains non-negligible. The transition density for 
transitions between nodeless orbits leads to a peak in the form factor at 
$q^2\!=\!2\Delta L/b^2\sim 2\Delta L\,A^{-1/3}$. As for {\piK} reactions, 
the minimum momentum transfer is large ($\sim 350$ MeV/c). This is beyond 
the peak of the form factor and means that all cross sections will fall 
with inreasing $q^2$ (reaction angle) and that high values of $\Delta L$ 
are favored. Thus, deeply-bound hypernuclear states (i.e. the ground state 
and states with the $\Lambda$ in low $L$ orbits) with both natural and 
unnatural parities may simultaneously appear, and provide a rich and new 
spectroscopy complementary to that from hadronic reactions. The high 
resolution allows extension of these studies to $sd$-shell states that 
could not be confirmed by low-resolution experiments or $\gamma$ 
spectroscopy.

 To compensate for the small electroproduction cross sections, high 
luminosity and forward spectrometer angles for both the scattered 
electrons and the reaction kaons are required. This creates a challenge 
to design an experiment with two large spectrometers essentially placed 
at zero degrees. Over the last decade, two independent hypernuclear 
programs in JLab Hall A~\cite{E94107} and Hall C~\cite{JLabexpts} 
have been developed and undertaken with encouraging 
results~\cite{Miyoshi03,Iodice07,Cusanno09,Nakamura13}.
The second and third phase Hall C experiments, E01-011 and E05-115, 
resulted in two new data sets, producing high resolution in the spectra 
of \lamb{7}{He}, \lamb{12}{B} and \lamb{28}{Al} and \lamb{7}{He}, 
\lamb{10}{Be}, \lamb{12}{B} and \lamb{52}{V}. This paper presents a 
combined analysis of the \lamb{12}{B} spectroscopy from the the Hall C 
program. The analysis of \lamb{7}{He} from the second phase experiment 
E01-011 has been previously published and papers describing the spectra 
from other hypernuclei are forthcoming. 

\section{\label{sec:design}Design of the HKS Experiments and Their
Apparatus}
\begin{figure}
\includegraphics[width=8cm]{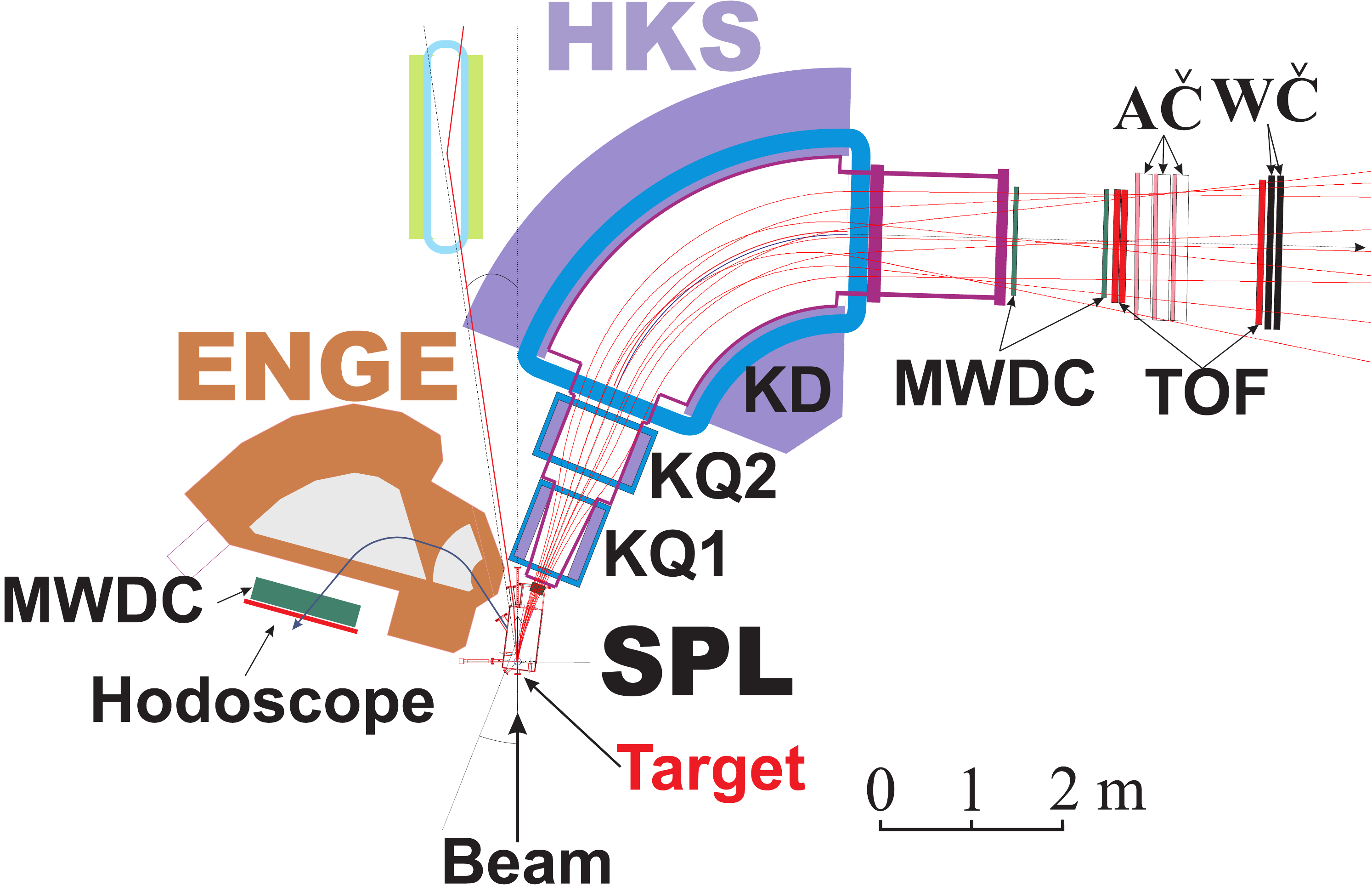}
\includegraphics[width=8cm]{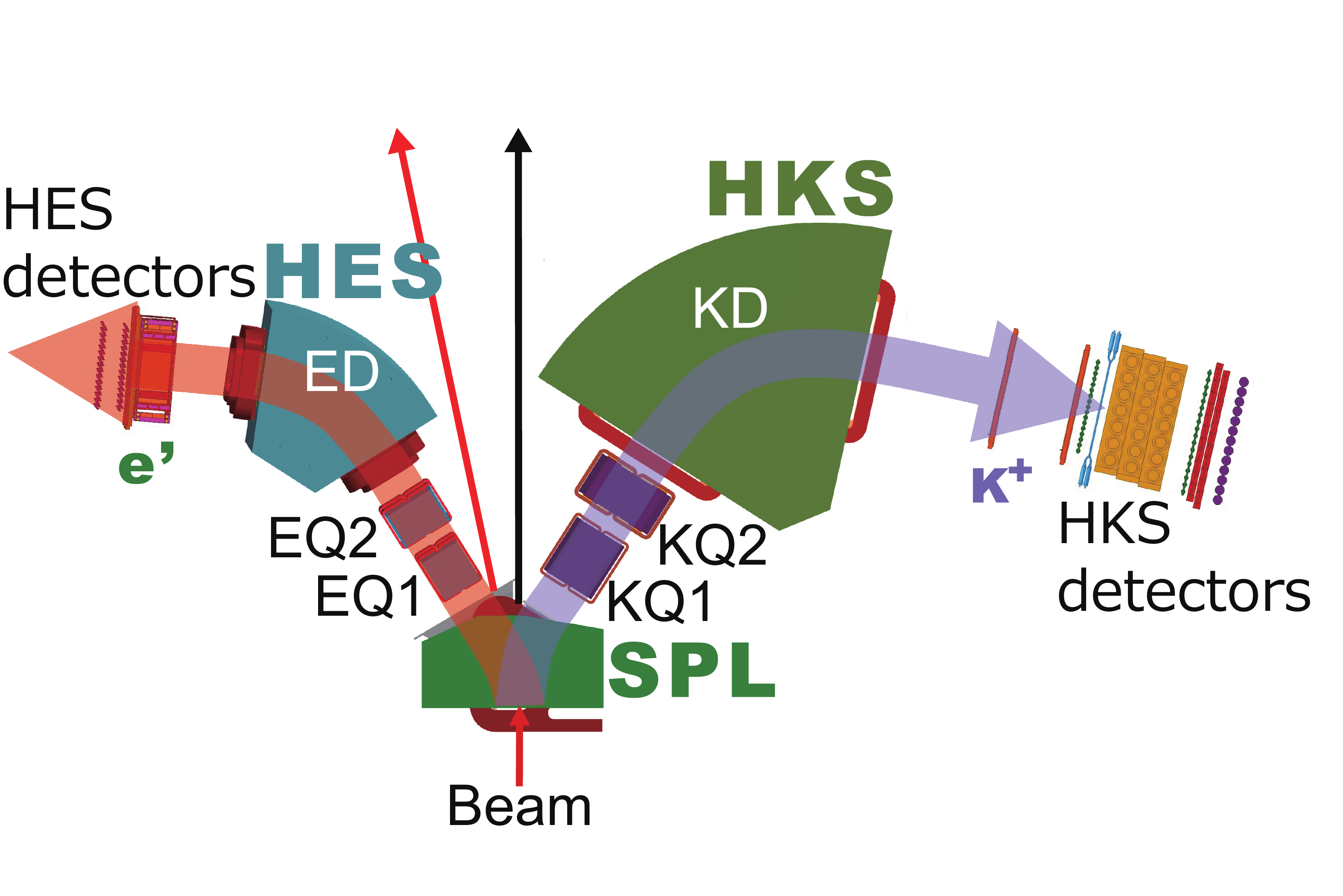}
\caption{\label{fig:figure1} (Color online) Schematic illustration of the
  experimental setup, technique and upgrades for the Hall C HKS
  hypernuclear spectroscopy experiments E01-011 (top) and E05-115
  (bottom).  }
\end{figure}

 The JLab Hall C HKS experiments, E01-011 and E05-115 are two, 
consecutive hypernuclear spectroscopy experiments (see schematic 
illustration in Fig.~\ref{fig:figure1}) which follow the first 
pioneering experiment (HNSS E89-009).  Upgrades and a new configuration 
were made to improve energy resolution, yield, and the use of a higher 
incident energy beam increased the virtual photon flux.  
Experiment E01-011 used a new, high resolution kaon spectrometer (HKS) 
having a short orbit and a large solid angle acceptance. An off 
scattering-plane geometry, the ``tilt method'', was applied to the electron 
spectrometer, the Enge Split-Pole spectrometer (Enge)~\cite{Enge79}. 
In E05-115, the previously used ``C'' type splitting magnet (SPL) and 
Enge were replaced by a new ``H'' type splitting magnet and a new high 
resolution electron spectrometer (HES) with a larger solid angle acceptance. 
The same ``tilt method'' which proved successful in E01-011 was also 
applied to the HES. The goal of this series of  upgrades was to improve 
precision and yield, in order to widen the spectroscopic studies beyond 
the $p$-shell.

\subsection{General Technique - A Common Splitter Magnet}
A charge-separation splitting magnet (SPL), common to both spectrometers,
is used by all the 
Hall C hypernuclear experiments in order to separate positive reaction 
kaons from the electrons.
The nuclear target under 
investigation is located at front effective field boundary (EFB) of the 
SPL which bends the oppositely charged particles ($e^-$ and $K^+$) away 
from the beam in the opposite directions.  This technique allows the 
spectrometers to be placed at forward angles close to the 
target.  As a result, the reaction particles are measured at very forward 
angles, with minimal path length for the short-lived kaons, and with increased
solid angle acceptance. All these are crucial factors which increase 
the yield.  However, a common SPL configuration also creates unavoidable 
challenges which are discussed in later sections.  

 The primary electron beam passes through the SPL and is deflected. At the 
high beam currents used in the two HKS experiments, the beam must be 
redirected to a high power beam dump in order to avoid serious radiation 
problems.  In E01-011, additional dipole magnets were installed downstream of the 
SPL to redirect the beam to the dump, while for E05-115 a pre-chicaned 
beam technique was applied to provide an incident beam angle which 
canceled the bending angle of the SPL.  Although the pre-chicane method 
requires careful tuning of the primary beam, it is significantly easier
than that of tuning the beam transportation after the SPL and provides
cleaner beam transport to the dump.

\subsection{Tilt Method}

 The extremely high electron singles rate in the electron spectrometer 
from Bremsstrahlung and M{\o}ller scattering presents another challenge 
at forward angles.  These background electrons are bent by the common 
SPL toward the electron spectrometer.  This problem limited the luminosity 
in the first experiment E89-009 (HNSS) to $0.4~\mu \text{A}$ on a 
$22~\text{mg}/\text{cm}^2$ thick C target, suppressing hypernuclear 
production while creating high accidental background.

The ``tilt method'' was developed for the latter two (HKS) experiments.  
The electron spectrometer (Enge for E01-011 and HES for E05-115) was 
tilted up, pivoting about a point approximately 43 cm upstream of the 
virtual target point, by an angle of $7^{\circ}$ off the plane as defined 
by the beam and the HKS momentum dispersion plane. This is equivalent to 
a rotation plus a shift of the spectrometer.  In such a configuration 
the scattered electrons at near zero degrees are blocked by the 
spectrometer yokes so that they lie outside the spectrometer acceptance. The 
rates for Bremsstrahlung and M{\o}ller scattering electrons decrease more 
rapidly with increases in scattering angle than does the virtual photon flux, 
especially when higher beam energies are used.  The tilt angle corresponds 
to a lower cut-off in the electron scattering angle of 
$\unsim 4.5^{\circ}$, a choice based on an optimization between 
the yield and the accidental background which could be accommodated by 
the experiments. Using this method, both E01-011 and E05-115 were able 
to increase the target thickness to $100~\text{mg}/\text{cm}^2$ and the 
beam current up to $40~\mu \text{A}$ while maintaining the electron 
singles rate at a level of approximately 3 MHz. This background was 
almost 100 times smaller than in the first experiment, E89-009,
improving the yield by more than an order of magnitude.

\subsection{Kinematics and Spectrometers}

\begin{table}[thb]
\caption{\label{tab:table1}
The basic kinematic and spectrometer parameters used for the
JLab Hall C experiments E01-011 and E05-115.}
\begin{ruledtabular}
\begin{tabular}{lcc}
\textbf{ITEMS} & \textbf{E01-011} & \textbf{E05-115} \\
\hline
\textbf{Beam energy} & 1.851 GeV & 2.344 GeV \\
Beam energy precision & $\leq 10^{-4}$  &$\leq 8\times 10^{-5}$ \\
\textbf{Electron spectrometer} & Enge & HES \\
Central $E^{\prime}$ & 0.351 GeV & 0.844 GeV\\
$\Delta E^{\prime}$&$\pm 25\%$ & $\pm 10.5\%$ \\
$E^{\prime}$ precision & $5\times10^{-4}$ & $2\times10^{-4}$ \\
$\theta_{e^{\prime}}$ minimum & $\unsim 4.5^{\circ}$ & $\unsim
4.5^{\circ}$ \\
$\Delta\Omega_{e^{\prime}}$ & 5.6 msr & 7.0 msr \\
\textbf{Average central $E_{\gamma}$}&1.5 GeV & 1.5 GeV \\
Average $Q^2$ & $\unsim 0.01~(\text{GeV}/c)^2 $& $\unsim
0.01~(\text{GeV}/c)^2$ \\
Average $W$ & $\unsim 1.90~\text{GeV}$ & $1.92~\text{GeV}$ \\
\textbf{Kaon spectrometer} & HKS & HKS \\
Central momentum $P_K$&$1.2~\text{GeV}/c$ & $1.2~\text{GeV}/c$ \\
$\Delta P_K$ & $\pm 12.5\%$ & $\pm 12.5\%$ \\
Precision & $\pm2\times10^{-4}$ & $\pm 2\times10^{-4}$ \\
$\theta_{eK}$ range & $1 - 13^{\circ}$ & $1 - 13^{\circ}$ \\
$\theta_{\gamma K}$ range & $0 - 12^{\circ}$ & $0 - 12^{\circ}$ \\
Average $\theta_{\gamma K}$ & $5.8^\circ$ &  $6.8^\circ$ \\
$\Delta\Omega_K$ & 16 msr & 8.5 msr \\
$K^+$ survival rate & $\unsim 30\%$ & $\unsim 27\%$ \\
\end{tabular}
\end{ruledtabular}
\end{table}

The basic parameters of the two experiments are listed in Table~\ref{tab:table1}.  
Although they used different beam energies, the virtual photon energy 
and its range were the same, so that the kaon spectrometer, HKS, did 
not need modification. In the sequence of upgrades, the substitution 
of the HKS for the original kaon spectrometer provided high kaon 
momentum resolution, while the further substitution of a new SPL and 
HES resulted in additional increase of yields.  Although this latter 
upgrade introduced a yield reduction from the HKS side due to the new 
SPL with a longer path, the larger solid angle acceptance from HES and 
more importantly, the higher beam energy which increased the total 
integrated virtual photon flux increased the yield by another factor 
of 4 for E05-115. 

 The beam energy was controlled by a high-frequency, fast-feedback, 
energy-lock developed at JLab. Furthermore, a Synchrotron Light 
Interferometer (SLI) was used in the Hall C beam line to measure 
and monitor beam stability and its variation in energy.  A more stringent
constraint on beam energy fluctuations was needed for E05-115 because of 
the higher beam energy. The chosen virtual photon energy of 
$E_{\gamma} \approx 1.5~\text{GeV}$ corresponds to approximately the 
maximum in the elementary $\Lambda$ photo-production cross section. This 
photon energy also optimizes the conditions for the HKS design with 
requirements for good kaon survival, large solid angle acceptance and
high resolution, and ease in kaon particle identification (PID). 
Note that at forward angles, the reaction $Q^2$ is sufficiently small so 
that virtual photons are almost real, and thus the \eepK\ cross 
section can be assumed to be approximately equal to the \gK\ 
differential cross section after integration over a virtual photon 
flux factor.  

\subsection{Detectors and Particle Identification}

\subsubsection{Detector System for the Electron Spectrometer}
 The detector system for the electron arm (both Enge for E01-011 and 
HES for E05-115) has tracking wire chambers to measure the focal plane 
parameters $(x, x^{\prime}, y, y^{\prime})$ and two segmented 
scintillation detector planes separated by 0.5 meters. The focal plane 
parameters, together with the point target position, are used to 
reconstruct the momentum and the scattering angle with optical 
reconstruction matrices obtained using the characteristics of the 
spectrometer.  The segmentation and the plane separation of the two 
scintillation detectors are designed to efficiently handle a high 
single-particle rate and to reject background particles originating
from outside the spectrometer acceptance. These two planes were also 
used to reconstruct the focal plane time reference which was then 
placed in coincidence with the $K^+$ in the HKS.  Since the rate of 
the scattered electrons is $10^4$ times larger than the sum of all 
the other negatively charged particles, particle identification (PID) 
in the electron spectrometer is not required. 

\subsubsection{Detector System for the HKS}

 The ``tilt method'' enabled a dramatic increase in the luminosity with 
respect to the first experiment. The luminosity increase also 
significantly increased the HKS singles rate.  Therefore, the HKS upgrades 
also included the installation of a sophisticated detector system.  This 
new system included the following:
\begin{enumerate}
\item Two sets of tracking wire chambers separated by 1.0 meters to 
provide precision measurement of the focal plane parameters;
\item Three layers of segmented scintillation detectors (two segmented 
in the momentum dispersion plane and one normal to the dispersion plane) 
separated by 1.75 meters. These provided a time-of-flight (TOF) 
measurement as well as providing a focal plane time reference when 
placed in coincidence with the detected electrons in the electron 
spectrometer;
\item Three layers of segmented Aerogel \v{C}erenkov (A\v{C}) detectors 
with n = 1.05 located between the second and third TOF planes which 
were used for $\pi^+$ and $e^+$ rejection;
\item Layers of segmented water \v{C}erenkov (W\v{C}) detectors with 
n = 1.33 installed behind the last TOF plane for proton rejection.
\end{enumerate}

"Bucking coils"~\cite{Gogami13} were used on each of the
photomultipliers in the \v{C}erenkov detectors. These coils made an active 
cancellation of the axial magnetic fringe field from the large HKS 
dipole and successfully restored the efficiency of these \v{C}erenkov 
detectors.

\subsubsection{Kaon Identification}
Several layers of \v{C}erenkov detectors, as described above, were arranged 
to provide a good background rejection power.  These detectors, which were
included in the trigger, maximized the kaon detection rate while limiting
the coincidence trigger rate such that the computer deadtime was kept below
10\%.
A Monte Carlo simulation was used 
to design a sophisticated online trigger scheme in an FPGA 
micro-processor~\cite{Okayasuthesis} which avoided accidental vetoes of 
$K^+$ from the high singles rate in the Aerogel \v{C}erenkov detectors 
and minimized background incident from outside of the spectrometer 
acceptance.

\begin{figure}
\includegraphics[width=8cm]{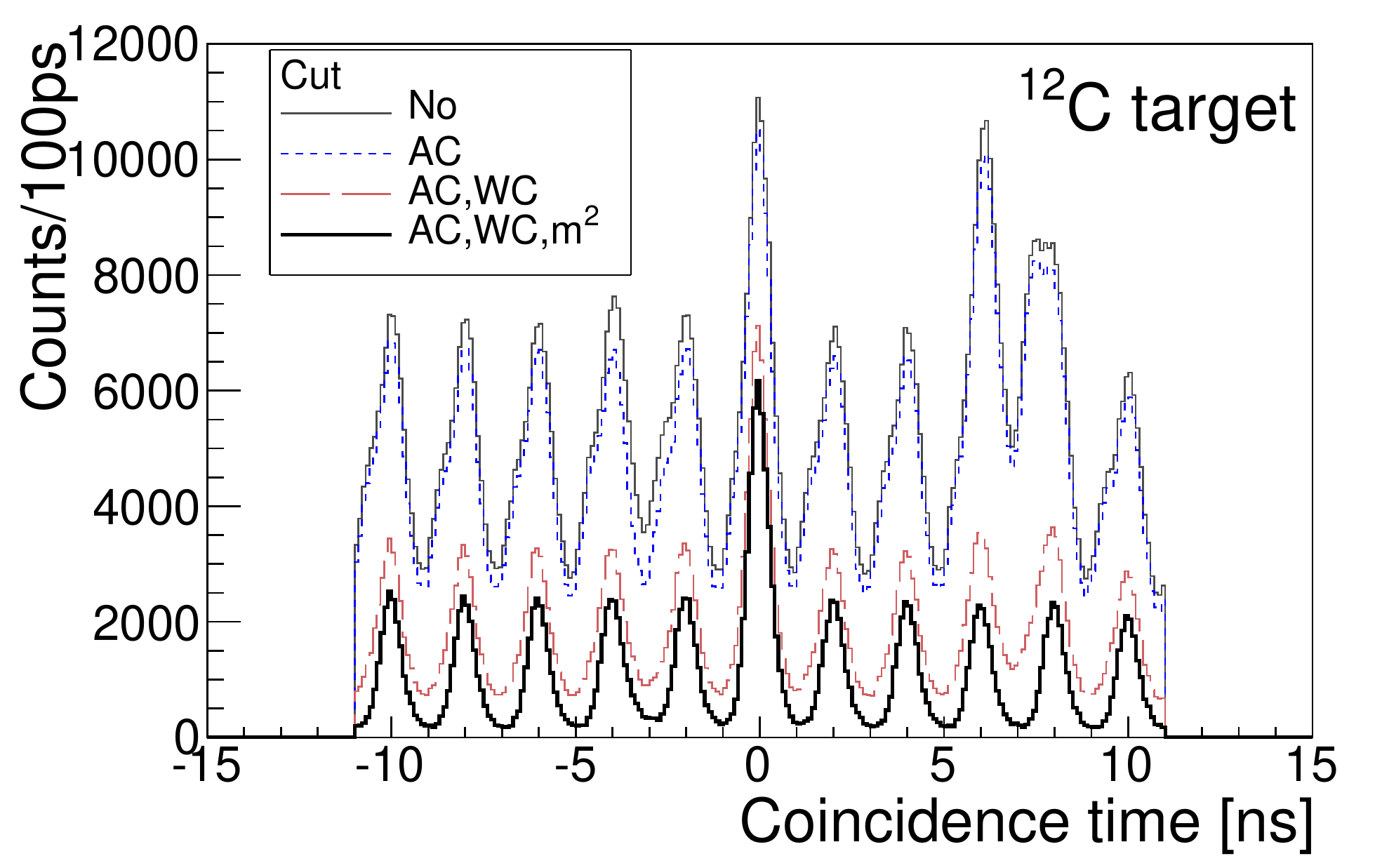}
\caption{\label{fig:figure2} (Color online) The coincidence time spectra 
with a sequence of the kaon identification cuts.}
\end{figure}

\begin{figure}
\includegraphics[width=8cm]{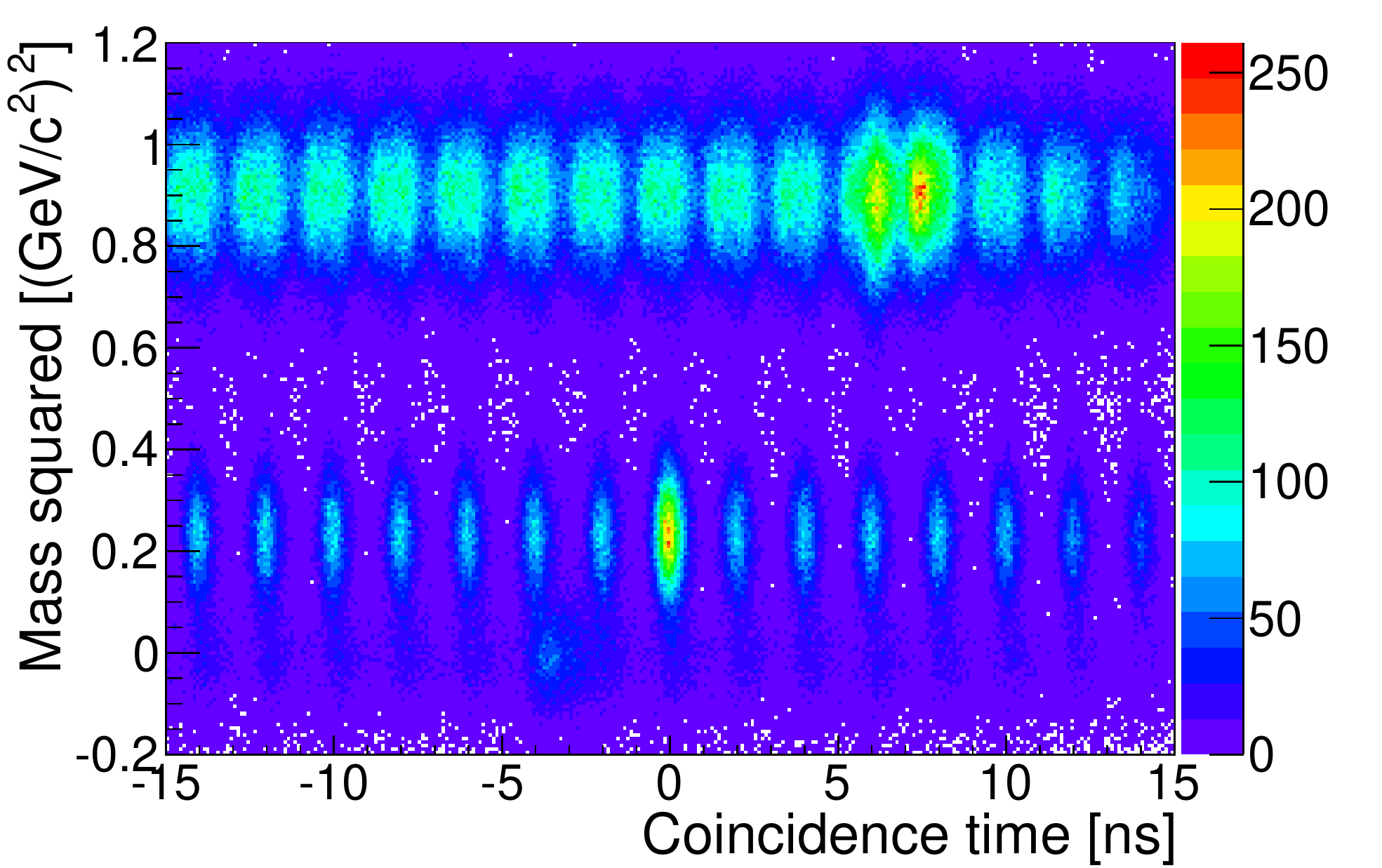}
\caption{\label{fig:figure3} (Color online) The 2-dimensional correlation 
between the derived mass squared and the coincidence time without the A\v{C} 
and W\v{C} cuts. After the A\v{C} and W\v{C} cuts, the three rows of pulse 
trains were clearly separated.}
\end{figure}

In the offline analysis, kaons were cleanly separated from the background 
particles ($e^+$, $\pi^+$, and $p$) by a combination of signals from the 
\v{C}erenkov detectors and the particle's mass squared ($m^2$) derived 
from measured velocity/TOF and momentum. Although the experimental 
conditions for the two HKS experiments were not identical (due to the 
various upgrades and technical changes), the basic technique and 
quality of the particle identification analyses were 
similar~\cite{Matsumurathesis,Rodriguezthesis,Sevathesis,Baturinthesis,Kawamathesis}.  
Figures~\ref{fig:figure2} and \ref{fig:figure3} 
demonstrate an example of the kaon identification as applied in the 
analysis of the data collected from the {\Ctwelve} target in the E05-115 
experiment.  Figure~\ref{fig:figure2} shows the coincidence time spectrum 
for the events detected by the electron and kaon spectrometers.  The full 
spectrometer path length correction was made in each spectrometer so that 
the time resolution was optimized, and the 2ns time interval of the 
CEBAF beam pulse separation can be clearly seen.  The distribution 
without ``Cut'' represents the minimum particle identification made at 
the trigger level.  The real coincidental pions and positrons
are located near -3 ns while the real protons are at $\unsim6.5$ ns in the 
plot.  This can also be seen in the 2-dimensional correlation between 
$m^2$ and coincidence time. After application of the A\v{C} and W\v{C} 
cuts and the 2-D gate on $m^2$, kaons in real or accidental coincidence 
with detected scattered electrons were cleanly separated from background.

\subsubsection{Accidental Background and Mixed Events Analysis}

 Due to the clean identification of kaons by the HKS detector system, the 
background in the reconstructed mass spectrum comes only from accidental 
coincidences.  The accidental coincidence level is not negligible because 
the electron singles rate is still high at the high luminosity which maximizes 
the production rate, and this background can only be removed by subtraction. 
However if the background shape is precisely measured, its contribution to 
the statistical error is small.  

 In order to precisely obtain the background shape and its height in the mass 
spectrum, a mixed event analysis was performed. Electron and kaon events from 
different accidental peaks (seen in Figs.~\ref{fig:figure2} and 
\ref{fig:figure3}) were randomly picked to create a mass spectrum using 
accidental timing which substantially increases its statistical accuracy. 
This spectrum was then scaled and used to subtract the accidental 
background.

\section{Kinematics and Optics Calibration}

 The development of the spectrometer calibration required the optimization 
of momentum and scattering angle reconstruction matrices. This was a 
complicated issue as the kinematics coupled the scattering angle and momentum 
measurement in each spectrometer.  As the beam passed through the common 
SPL it was impossible to use an elastically scattered, monochromatic 
beam to separately obtain a momentum for each scattered particle.  In addition, 
both the elastically scattered and primary beam electrons passed through the 
focal plane of the spectrometers when at forward angles. 

 Although a Sieve Slit (SS) collimator is a device commonly used with
magnetic spectrometers to extract the momentum and angle transfer matrices of a 
spectrometer, it is difficult to use in the geometry of these experiments.
An SS collimator
is a thick plate with arrays of well positioned small holes which 
is mounted between a spectrometer and an experimental target.  When 
a point beam ($\unsim100~\mu\text{m}$ diameter) scatters from a target, the 
center of each small hole defines a uniquely known position and angular 
coordinate. Events from each hole form a pattern on the focal plane such that 
reconstruction matrices can be mathematically fitted. However the SS placed in 
front of each spectrometer in the HKS geometry was behind the common SPL 
which introduced momentum and angle correlations.  Thus events from a given hole 
cannot select a small kinematic volume with a unique angle and momentum. 
Thus special techniques are required which use events from the {\eepK} reaction 
on targets with well-known masses.  Although such calibrations are possible, 
they are difficult and time consuming.

\subsection{\label{sec:magfieldinterfere}Magnetic Field Interference and Corrections}

 In order to overcome the difficulties mentioned above, extensive GEANT4 
Monte Carlo simulations were run with both measured 3-D magnetic field maps and 
a fields maps from a finite element calculation by Opera-3D (TOSCA).  These 
simulations were used to study the momentum, angular resolution, and acceptance 
of the spectrometers, and used to evaluate calibration methods, procedures and 
uncertainties using the simulated {\eepK} reaction with well-known masses.  
They produced correlations between the focal plane parameters from the 
simulated SS events generating initial backward reconstruction matrices. 
Finally, the matrices obtained from simulation were optimized using events 
selected from real data. 

\begin{figure}
\includegraphics[width=8cm]{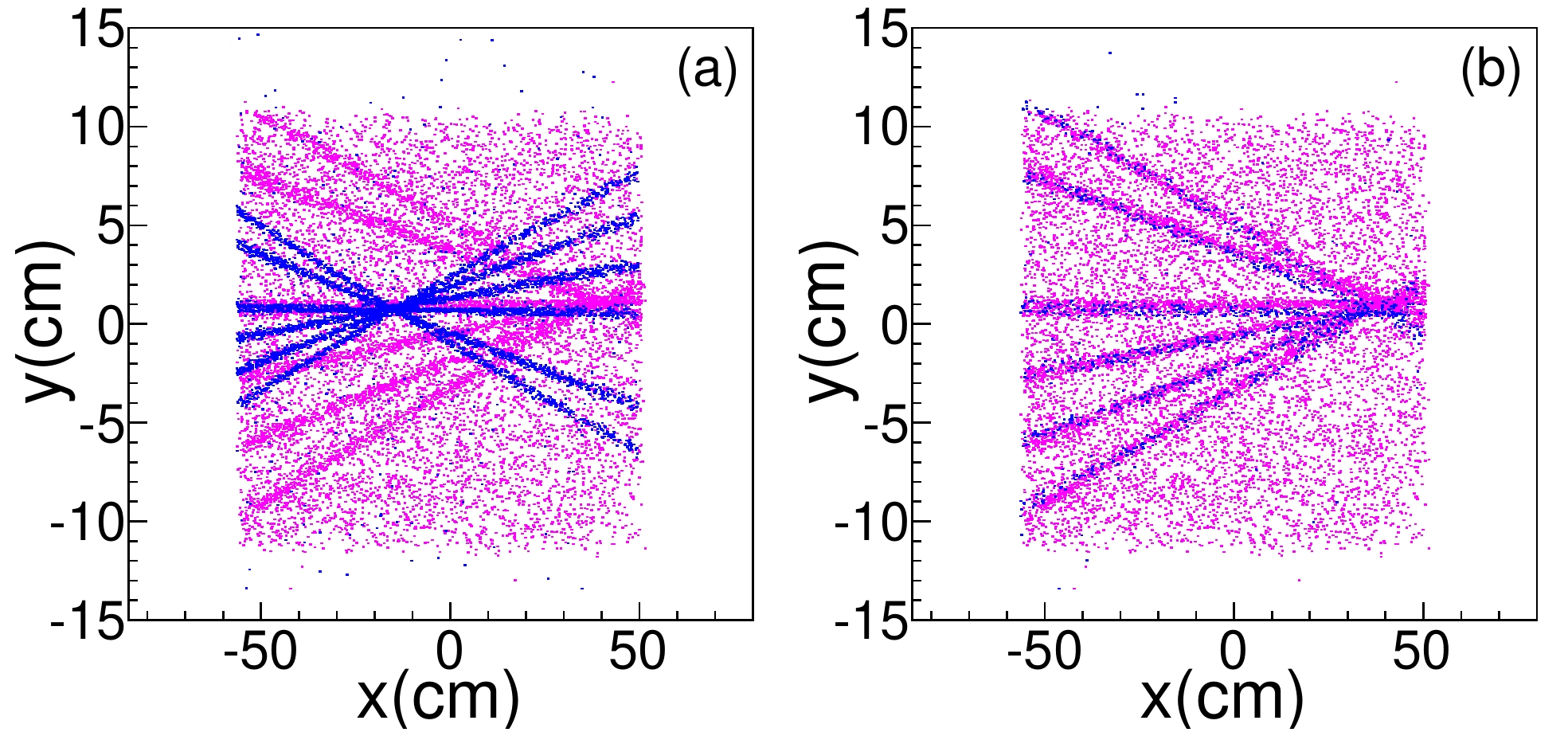}
\caption{\label{fig:figure4} (Color online) The HKS focal plane $Y$ vs $X$
  correlations for the events selected from the $7^\text{th}$ column
  of the SS holes.  Blue - simulated data and Magenta - real data.  
(a) before field correction and (b) after field correction.}
\end{figure}

 A problem appeared when comparing focal plane parameter correlations 
(such as $Y$ vs $X$) between the simulated and the real SS events from the 
E05-115 experiment (2009).  Reasonable agreement was seen only for the 
events coming from the column furthermost from the beam centroid (i.e. away 
from HES).  The disagreement increased as the SS column approached the 
beam centroid, while the symmetry remained in the non-dispersive plane 
($Y$-$Z$).  For example, Fig.~\ref{fig:figure4} (a) shows a comparison 
for the events selected from the 7th column of the SS holes for the HKS 
spectrometer.  This column of SS was on the side toward to HES where a 
line of events came from a specific hole in that column. Similar behavior 
was also found for the HES spectrometer, i.e. the disagreement started and 
enlarged as the selected column came closer to the HKS spectrometer. Due to 
the ``tilt'' of the HES, the symmetry about the central angle was not 
expected and disagreement was found in both ($X$-$Z$ and $Y$-$Z$) planes.

\begin{figure}
\includegraphics[width=8cm]{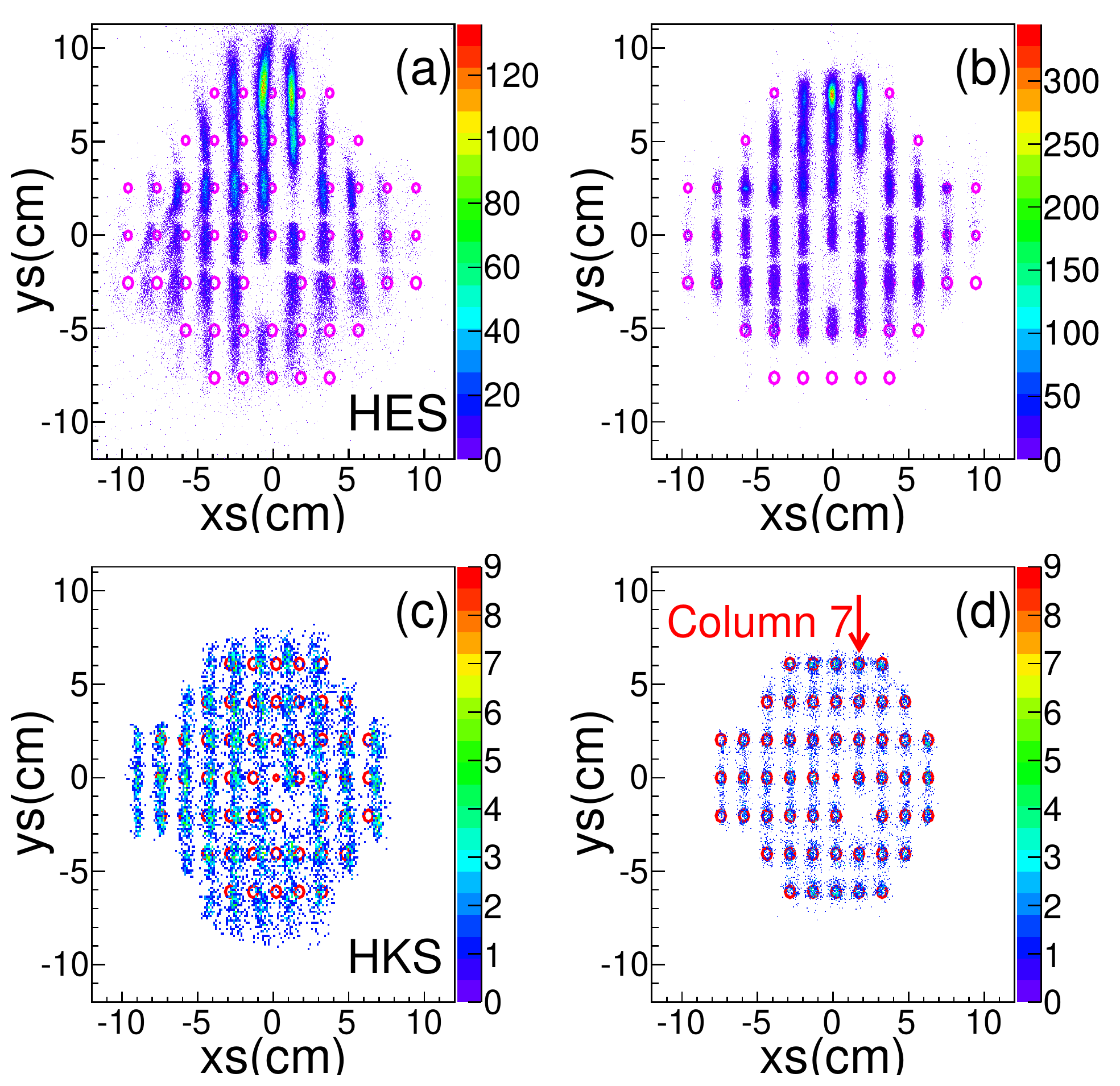}
\caption{\label{fig:figure5} (Color online) The reconstructed real HES and 
HKS SS events at the SS plate in comparison to the actual geometry of the
SS holes.  (a, c)before field correction and (b, d) after field
correction.}
\end{figure}

 The disagreement indicated that the reconstruction matrices obtained from 
simulations were not sufficiently close to the correct ones to use as initial 
values in a perturbative development of the real optics. 
Figure~\ref{fig:figure5} (a) and (c) show the comparisons of the reconstructed 
real SS events at the SS plates to the actual geometry of the SS plates for 
HES and HKS, respectively. The disagreements and asymmetry are obvious and 
significant. The particle density variation in the case of the HES was strongly 
dependent on the angle of the scattered electrons. This problem was studied and 
found to be a consequence of field interference between the SPL magnet and the 
front magnetic elements of the spectrometers. Due to the asymmetry in the relative 
geometry between SPL and spectrometers, a TOSCA calculation could not combine 
independent measurements of the spectrometer fields. 

 The problem was resolved by the addition of field corrections to the 3-D 
field map used in the GEANT4 simulation.  The three dimensional field 
corrections are assumed to have coordinate dependencies described by polynomial 
functions.  These were applied to the field for each appropriate element and 
the coefficients of the polynomials were then optimized to minimize the 
variation between the simulated events from each SS hole in comparison to the 
real events.  Fig.~\ref{fig:figure4} (b) demonstrates the results when 
comparing the simulated and real SS events from the $7^\text{th}$ column of SS 
holes for HKS. The systematic tuning of the polynomials resulted in uniform 
agreement over the full kinematic space $\Delta P \Delta \Omega$ for both HES 
and HKS.  This provided optical transfer matrices sufficiently close to the 
correct values to provide initial starting values for further optimization as 
demonstrated by Fig.~\ref{fig:figure5} (b) for HES and (d) for HKS.

As the experimental configuration was similar, the same problem was 
confirmed to exist in the E01-011 (2005) experiment. Therefor the E01-011 data 
were reanalyzed with the same technique.
An independent analysis of 
the E01-011 data reached the same level of agreement. 

\begin{figure}
\includegraphics[width=8cm]{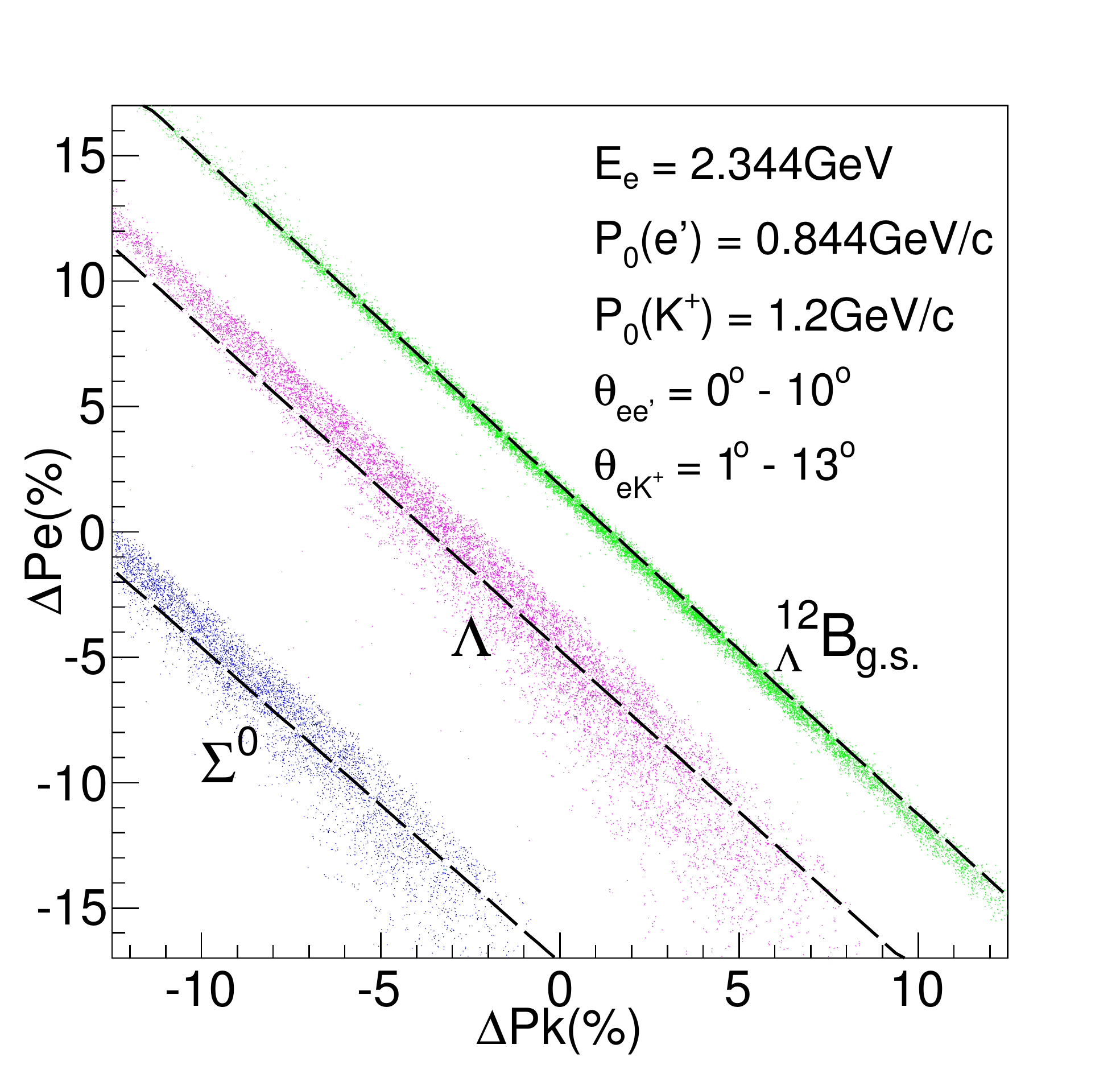}
\caption{\label{fig:figure6} (Color online) The mass correlations of free 
$\Lambda$ and $\Sigma^0$ from protons, and the ground state of {\BLtwelve} from
{\Ctwelve}, from the {\eepK} reaction.}
\end{figure}

\subsection{Kinematics Calibration}

 The large momentum acceptances of both the electron (Enge and 
HES) and kaon (HKS) spectrometers can capture, in a single setting,
events from free $\Lambda$, and free $\Sigma^0$ production from protons in a {\CHtwo} 
target, and hypernuclear events from different nuclear targets. 
Figure~\ref{fig:figure6} illustrates the mass correlation between the momenta 
of electrons and kaons from the {\eepK} reaction for production of $\Lambda$ 
and $\Sigma^0$ from hydrogen in a {\CHtwo} target, and the ground state of 
\lamb{12}{B} from a {\Ctwelve} target. The correlations are the same for both 
E01-011 and E05-115. The dispersion of the events from the locus line is due 
the angular acceptance of the spectrometers. The dashed lines show the events 
with central angles. Since the masses of free $\Lambda$ and $\Sigma^0$ are light, 
they have strong angular dependencies.  In contrast, this dispersion becomes 
much smaller for heavy systems, such as hypernuclei, as seen for the 
\lamb{12}{B} distribution in Fig.~\ref{fig:figure6} from E05-115.

\begin{figure}
\includegraphics[width=8cm]{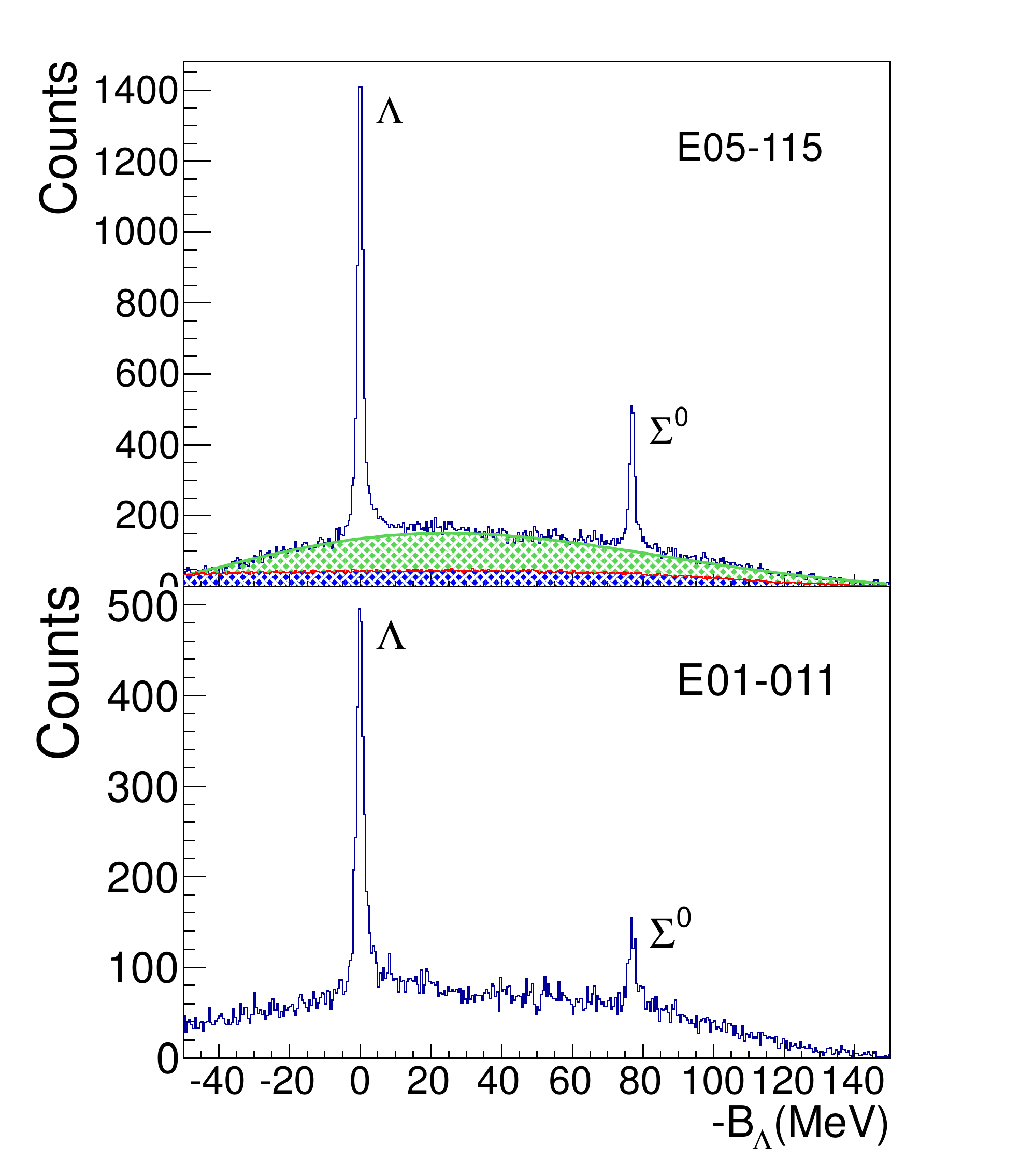}
\caption{\label{fig:figure7} (Color online) Spectroscopy of free $\Lambda$ 
and $\Sigma^0$ by the $p\eepK\Lambda$ reaction from the {\CHtwo} target. 
The mass is presented in terms of $\Lambda$ binding energy.}
\end{figure}

\begin{table*}[bht]
\caption{\label{tab:table2} The reconstructed mass and separation of
  $\Lambda$ and $\Sigma^0$, in $\text{MeV}/c^2$, from the two experiments.  
The PDG values of   $M_{\Lambda} = 1115.683\pm0.006$ and 
$M_{\Sigma^0} = 1192.642\pm   0.024$ $\text{MeV}/c^2$ are used.}
\begin{ruledtabular}
\begin{tabular}{cccccc}
 & $B_{\Lambda}(\Lambda)$ & Width (FWHM)&$B_{\Lambda}(\Sigma^0)$
  & Width (FWHM) & $\Delta M(\Sigma^0-\Lambda)$ \\
\hline
E05-115 & $-0.030\pm0.014$ & $1.946\pm0.033$ & 
$76.945\pm0.028$&$1.849\pm0.071$ & $76.965\pm0.031$ \\
E01-011 & $0.014\pm0.033$ & $2.583\pm0.079$ & $77.001\pm0.094$ &
$2.672\pm0.247$ & $76.987\pm0.259$ \\
\end{tabular}
\end{ruledtabular}
\end{table*}

  Simultaneous production of free $\Lambda$, $\Sigma^0$ and hypernuclei is 
a major advantage of the HKS experiments. The masses of $\Lambda$ and 
$\Sigma^0$ are sufficiently well known and their mass separation 
($76.92~\text{MeV}/c^2$) is large.  This allows precise kinematic calibration 
of the spectra and an absolute mass scale calibration. Figure~\ref{fig:figure7} 
shows the final mass spectroscopy of $\Lambda$ and $\Sigma^0$ in terms of 
$\Lambda$ binding energy from both the E01-011 and E05-115 experiments. The 
background includes accidentals and the \lamb{12}{B} quasi-free production 
from {\Ctwelve} in {\CHtwo}. The spectra are analyzed using $p\eepK\Lambda$ 
kinematics.  The accidental background shape can be determined precisely by a
mixed event analysis and the quasi-free background shape is experimentally 
obtained from carbon target data which is analyzed with $p\eepK\Lambda$ kinematics. 
Therefore, the background shape is almost completely understood for the 
{\CHtwo} data.

 Table~\ref{tab:table2} lists the reconstructed masses of $\Lambda$ and 
$\Sigma^0$. The kinematic calibration was undertaken in concert with other 
calibrations and optical optimizations which will be discussed in the later 
sections. The uncertainty in the calibrated mass scale contributes to the 
systematic uncertainty in the absolute hypernuclear mass scale. The total 
systematic uncertainty of $\Lambda$ and $\Sigma^0$ masses includes the 
statistical uncertainty as listed in Table~\ref{tab:table2} and systematic
uncertainties due to the radiative tails and background/peak fitting functions. 
The radiative tail was studied with the Hall C SIMC code~\cite{SIMCref} and a 
correction was applied to minimize the mass offset residuals. The contribution 
from this calibration to the overall systematic uncertainty in the absolute 
binding energy of hypernuclei is found to be $\pm 27~\text{keV}$ and 
$\pm 43~\text{keV}$ for E05-115 and E01-011, respectively. However, this 
uncertainty is not present in the excitation energy spectrum with respect to 
the ground state (or in the energy separation between states).  The mass 
separation uncertainty is found to be less than $\pm 70~\text{keV}$ over the 
$\unsim 77~\text{MeV}/c^2$ mass range between $\Lambda$ and $\Sigma^0$. The 
excitation energy uncertainty is less than $\pm 10~\text{keV}$ for both 
experiments in an approximate 10 MeV range in excitation energy above the 
ground state.

\subsection{Optical Matrix Optimization}
 For a point beam on target with stabilized position, the target coordinate
set is $(X=0, X^{\prime}, Y=0, Y^{\prime}, L=0, \delta)_t$.
$X^{\prime}$ and $Y^{\prime}$ are the angles in and off the momentum dispersion
planes with respect to the spectrometer optical Z axis, respectively and are 
related to the scattering angle of the detected particle. L=0 is the reference 
point of the trajectory path length and $\delta$ is the percentage momentum
offset for the detected particle relative to the central momentum setpoint for the 
spectrometer. Correspondingly, at focal plane (FP) of the spectrometer,
the particle's 
coordinate set is $(X, X^{\prime}, Y, Y^{\prime}, L, \delta)_{\text{FP}}$,
in which $X$, $X^{\prime}$, $Y$ and $Y^{\prime}$ are measured quantities.
Mathematically,
the FP coordinates are the matrix-vector product of the spectrometer
optical transportation matrix and the target coordinate vector. The field 
interference correction work done by the GEANT simulations described in 
Section~\ref{sec:magfieldinterfere} serves to find a transport matrix
that is close to the real optics.

  Each of the four unknowns, $X^{\prime}_t$, $Y^{\prime}_t$, $\delta$ and 
$L$ is then obtained separately from the product of a reconstruction optical matrix 
and the FP coordinate vector with the others assumed known. The initial matrix 
is sufficient for the path length $L$ reconstruction which provided the full path
length correction to the coincidence time between $e^{\prime}$ and $K^+$.
As its
contribution to the precision of the momentum and angle reconstruction is 
negligible, the path length need not be further optimized. For 
each spectrometer, there are three optical reconstruction matrices, for
$X^{\prime}_t$, $Y^{\prime}_t$ and $\delta$, that must be optimized as
they are crucial to achieving the best resolution.
The optimization of these matrices is another challenge to these experiments.
As mentioned previously, the common splitter prevents single spectrometer 
calibration using two-body scattering.   Thus, the matrices can only 
be optimized by using well defined physical events from an \eepK\ reaction. The 
difficulty is that the six matrices (three for each spectrometer) are 
coupled through the reaction kinematics so that matrices from
different spectrometers
affect each other. On the other hand, small errors can compensate each
other so that
the derived invariant mass and scattering angles are somewhat insensitive
to these errors.  To resolve the complications of this coupling
special techniques and optimization procedures were 
developed, aided by extensive simulation studies.
There are $\unsim 1300$ matrix parameters in the six matrices
which include terms from order ($0^{\text{th}}$ to 
$6^{\text{th}}$).  Due to the kinematic coupling and compensation effect
between the two spectrometers, the six matrices are separated into two groups:
the momentum reconstruction matrices (one matrix from each spectrometer) and 
the angle reconstruction matrices (two matrices from each spectrometer).
Each group is optimized separately and the improvement of each group allows
the other group to be further improved.

\subsubsection{Optimization by the $\Lambda$ and $\Sigma^0$ Productions}
 Both $\Lambda$ and $\Sigma^0$ produced by the {\eepK} reaction were used to 
optimize the momentum and angular reconstruction matrices using a standard 
least-$\chi^2$ minimization method. For each event selected from 
the peak of $\Lambda$ (or $\Sigma^0$), the difference between its reconstructed 
mass and the corresponding reference mass (the PDG value of $\Lambda$ or 
$\Sigma^0$ as given in the caption of Table~\ref{tab:table2}) was used in
the computation of $\chi^2$.
$\chi^2$ was then minimized by varying
the matrix parameters.  Events were selected from the $\Lambda$ and $\Sigma^0$ 
peaks within a width of $\unsim\pm 1.5\sigma$ about the mean of the peak 
values. A width limit was applied because background events were unavoidably 
included and widening the gate decreased the signal to background ratio 
reducing the sensitivity of the fit.  The minimization was iterated 
by alternately optimizing the momentum and angle matrices.

\begin{figure}
\includegraphics[width=8cm]{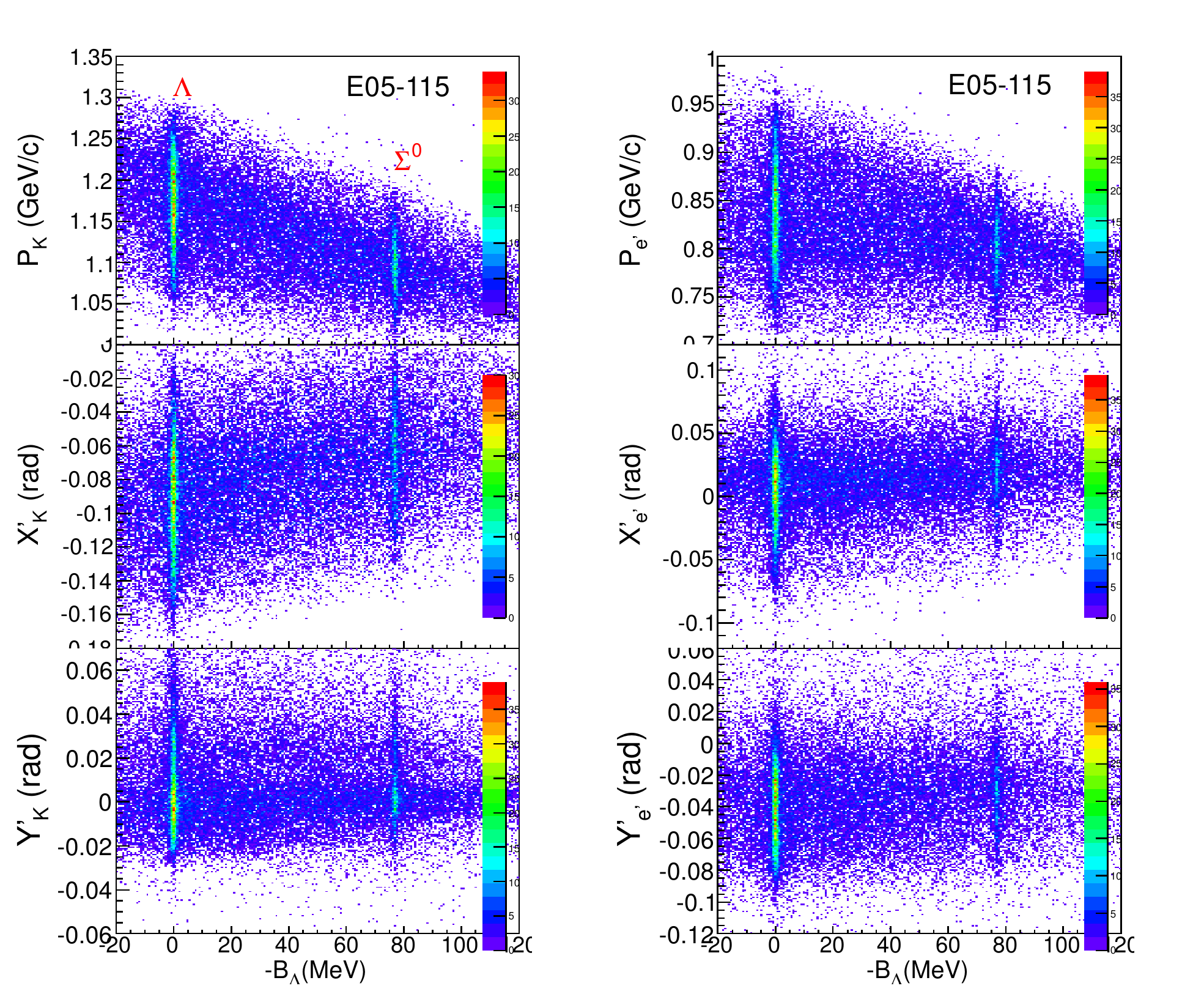}
\caption{\label{fig:figure8} (Color online) Correlations of the reconstructed
  kinematics parameters (P, $X^{\prime}$ and $Y^{\prime}$ at the target) from 
  both spectrometers to the calculated invariant masses of $\Lambda$ and 
  $\Sigma^0$ from the E05-115 analysis.  The E01-011 analysis shows similar
  features except with lower statistics.}
\end{figure}

Figure~\ref{fig:figure8} shows the correlations between the reconstructed 
invariant masses of $\Lambda$ and $\Sigma^0$ in terms of the $\Lambda$
binding energy and the reconstructed parameters (absolute momentum, $P$, in-plane and
off-plane angles at target, $X^{\prime}$ and $Y^{\prime}$) from which the 
invariant masses were calculated. The six correlations correspond to the six 
reconstruction matrices (three from each spectrometer). This figure
verifies the quality of the optimized matrices, showing that 
the calculated invariant mass is independent of the reconstructed kinematics 
parameters ($P$, $X^{\prime}$ and $Y^{\prime}$ at target).
The local 
mass spread is minimized to ($<100~\text{keV}/c^2$). Thus the quality 
and precision of the optimized optics are uniform in the complete kinematic 
space, so that the energy resolution as well as the excitation energy scale 
is essentially uniform.  

\subsubsection{Beam Position Correction}

To prolong the lifetime of the {\CHtwo} target which can be damaged by an 
intense primary electron beam, a fast raster moved the beam over the target 
at 20 kHz.  Thus the point electron beam ($\unsim 100~\mu\text{m}$) was 
distributed over an area $\unsim 8\times 8~\text{mm}^2$ for E01-011 and 
$\unsim 3\times 6~\text{mm}^2$ for E05-115.  The effect on the spectrum is 
the same for both experiments, yielding both momentum and angles offsets
that depend on the raster size and spectrometer optics.
GEANT4 studies with a realistic 
field map show the effect can be eliminated by a correction to the focal 
plane parameters (i.e. the measured $X$, $X^{\prime}$, $Y$ and $Y^{\prime}$ at
the FP), allowing reconstruction matrices for point target to be used. The 
correction was realized event by event by using a set of matrices which 
calculated the correction from the defined raster phase angle and amplitude 
in both the $X$ and $Y$ directions. The least-$\chi^2$ method was 
used to optimize the phase angle and amplitude as they were not precisely 
measured. Similarly, the correction matrices were also optimized using the
same method as for the optimization of the reconstruction matrices. This procedure
was repeated at different stages in the progress of the optimization 
of the momentum and angle matrices.  Removing this contribution of rastered
beam position ensures
that the optimized optical reconstruction matrices using the $\Lambda$ and 
$\Sigma^0$ events are valid for a point target, as the targets used in 
producing hypernuclei were used with unrastered beam.

\subsubsection{Target Straggling and Kinematics Alignment}
\begin{figure}
\includegraphics[width=8cm]{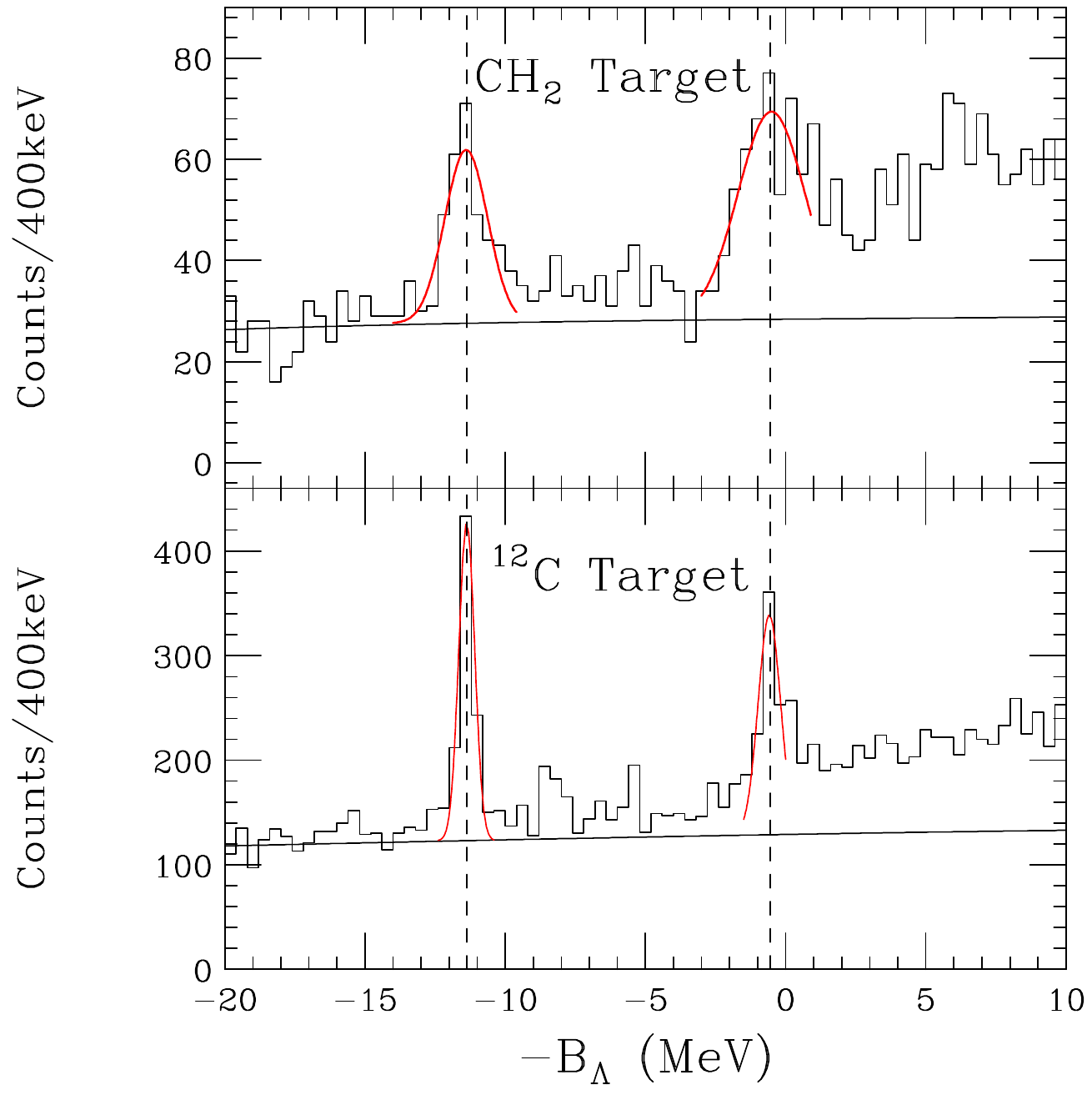}
\caption{\label{fig:figure9} (Color online) Spectroscopy of \lamb{12}{B} 
obtained by the $\Ctwelve\eepK$ reaction from the {\CHtwo} and {\Ctwelve}
targets from experiment E05-115.  The same alignment procedure was
followed for E01-011.}
\end{figure}

Since the $\Lambda$ and $\Sigma^0$ events from a {\CHtwo} target were used for 
both kinematic calibration and optimization of the reconstruction matrices, the 
scattering kinematics for each event must be known accurately.
One issue which arises is that the thickness of the 
{\CHtwo} target is not accurately known. Incorrect mean target straggling 
corrections can result in an incorrect optimization of the reconstruction matrices 
and thus affect the energy resolution for mass spectroscopy
of hypernuclei.
Therefore, a {\Ctwelve} target with a well known foil 
thickness was used to obtain an effective thickness.  Target straggling and 
energy corrections as function of the thickness were studied with the Hall C SIMC 
code and a GEANT4 simulation.  Straggling corrections were applied to the {\Ctwelve}
target data in order to obtain the \lamb{12}{B} spectrum.  Similarly, events 
from the {\CHtwo} target were also analyzed with $\Ctwelve\eepK$ kinematics 
to generate a \lamb{12}{B} spectrum formed from events from the {\Ctwelve} component of 
the {\CHtwo} target. Although statistics were lower, the $s$-shell ground and 
$p$-shell substitutional state peaks are well recognized.  Corrections 
corresponding to various {\CHtwo} target thicknesses were scanned to find the 
best simultaneous alignment of both the ground and $p$-shell states between the 
{\CHtwo} and {\Ctwelve} spectra.  Uncertainty in the alignment is dominated by the
statistical uncertainties of the two peaks.  This contributes to the systematic 
uncertainty in the determination of the binding energy with the defined 
kinematics and optics. Figure~\ref{fig:figure9} shows the alignment of the 
two spectra from E05-115 data. The $s$- and $p$-shell peaks are aligned within 
10 keV, and the statistical uncertainties of the $s$- and $p$-shell peaks 
from the {\Ctwelve} target are $\pm 19~\text{keV}$ and $\pm 33~\text{keV}$ keV, 
respectively.  Due to low statistics, the uncertainties for the {\CHtwo} 
target are $\pm 130~\text{keV}$ and $\pm 190~\text{keV}$. To reduce the 
overall alignment uncertainty, two independent single peak alignments were
done by the $s$- and $p$-shell peaks separately and an average was done
taking into account the statistical uncertainties of each peak. The overall 
alignment uncertainty is then found to be $\unsim \pm 90~\text{keV}$ and 
$\unsim \pm 140~\text{keV}$ for E05-115 and E01-011, respectively.

 This correction for the {\CHtwo} target thickness enables the use of the 
$\Lambda$ and $\Sigma^0$ peak positions for the kinematics calibration. It 
ensures that the optimization of the reconstruction matrices is done in one unified 
kinematics so that the matrices are applicable to data from all production 
targets used in the experiment. All other production targets were then 
separately optimized for their own target straggling corrections using 
this unified kinematics. The residual relative alignment error affects only 
the absolute binding energy but does not affect the excitation energy which 
is measured relative to ground state.

The above procedure was iterated several time, alternating with other 
optimization procedures until the derived {\CHtwo} thickness was stable.
Iteration was needed because of the kinematic coupling in the events 
produced by the $\eepK$ reaction. Improvement from each aspect of optimization
allowed the other parts to be further improved.

\subsubsection{Optimization Involving Events from \lamb{12}{B}}

 Events selected from the peaks of $\Lambda$ and $\Sigma^0$ cannot be used 
alone to fully optimize the momentum and angular reconstruction matrices. 
This is because the reactions on protons which produce these recoil
particles result 
in a large recoil kinetic energies due to the light masses involved. Monte Carlo 
studies demonstrated that these spectra are almost equally sensitive to the 
uncertainties in recoil momentum and the angular matrices.  On the other hand, because 
of their heavier masses, recoil energies in the production of hypernuclei are small, and 
thus the widths of the hypernuclear states depends almost entirely on the 
uncertainty in the momentum matrix.  For example, the width of \lamb{12}{B} 
states has approximately 40 times smaller angular dependence than does the $\Lambda$ peak. 
To mitigate this problem, events from peaks of well-defined heavy mass 
targets must be simultaneously used together with the events producing 
$\Lambda$ and $\Sigma^0$ recoils. Since the hypernuclear peak widths 
essentially depend only on the uncertainty in the momentum matrices, such 
events can be used to insert a known functional dependence of the momentum 
vs.~angle correlation, and optimize the momentum matrix with less influence
from uncertainties in the angle matrices. Once an improved momentum matrix 
is obtained, it is then used to optimize the angular reconstruction matrices 
with only $\Lambda$ and $\Sigma^0$ events. Inclusion of $\Lambda$ and 
$\Sigma^0$ data in the momentum reconstruction matrix optimization with a small 
weight in the overall $\chi^2$ definition is necessary to ensure uniform energy 
resolution over the large kinematic space. This procedure is iterated to 
convergence and is combined with other corrections and optimizations discussed
previously to form a complete optimization cycle.

 The $\Ctwelve\eepK\lamb{12}{B}$ reaction has a large cross section, has been 
previously studied in several electroproduction experiments, and the ground 
state has been observed in emulsion data.  In addition, an extensive knowledge 
of the states of its isospin mirror partner, \lamb{12}{C}, exists.
The \lamb{12}{B} ground state and the strongly excited $p$-shell peak are 
suitable calibration states for momentum matrix optimization and sufficient 
statistics can be reached within a relatively short beam time. Therefore, 
events from these states were selected for this optimization procedure. The 
momentum reconstruction matrix is required to fit the mean kinematics as 
defined by the mass of $\Lambda$ and $\Sigma^0$ using the PDG values (see
in caption of Table~\ref{tab:table2}).  The masses of the two experimentally 
measured \lamb{12}{B} states were then allowed to vary, and a statistical mean 
width for each peak was used to define $\chi^2$, together with that from 
$\Lambda$ and $\Sigma^0$ as mentioned above.  A minimization of this width, 
keeping the energy scale fixed (locked by the $\Lambda$ and $\Sigma^0$ masses), was 
obtained by minimization of the overall $\chi^2$. It was found that the mean
mass of these two \lamb{12}{B} states became stable within a few keV once a 
width of $\unsim 1.0~\text{MeV}$ FWHM was reached in the progression of 
optimization iteration.

\subsubsection{Blind Simulation Analysis and Systematic Error from 
Matrix Optimization}
 A blind analysis to a simulated data set was carried out to evaluate the 
systematic error generated by the matrix optimization processes. This method 
was also used to study the contribution from each individual source, to the 
accuracy of the focal plane parameters, the target thickness and energy 
straggling, the beam position raster size, and the angle and momentum 
uncertainties from the optimized reconstruction matrices.  Thus a Monte Carlo 
simulation was used to generate calibration data for photoproduction on 
protons producing $\Lambda$ and $\Sigma^0$, and hypernuclear states from a 
{\Ctwelve} target. The quantities of simulated data corresponded to experimental 
quantities. The masses of the hidden hypernuclear states were then extracted 
using the same optimization procedures as in the experimental data. The results 
of this study allowed an estimate of the error in how well a mass could be 
determined by the analysis techniques described above.  The study concluded 
that the optics matrix optimization contributes uniformly a 
$< \pm 50~\text{keV}$ systematic uncertainty to the determined mass for all 
studied hypernuclei.  Energy resolution (i.e. the peak width determination) 
for each production target was also found to be uniform (less than a few keV 
fluctuations) within the applicable excitation energy range. The resolution
varies only between different targets due to target thickness and mass 
differences. Both kinematic calibrations and the optical optimizations
contribute a systematic uncertainty which results a small constant error 
in determination of binding energy of all states but not the relative energies 
between states.

\begin{figure}
\includegraphics[width=9cm]{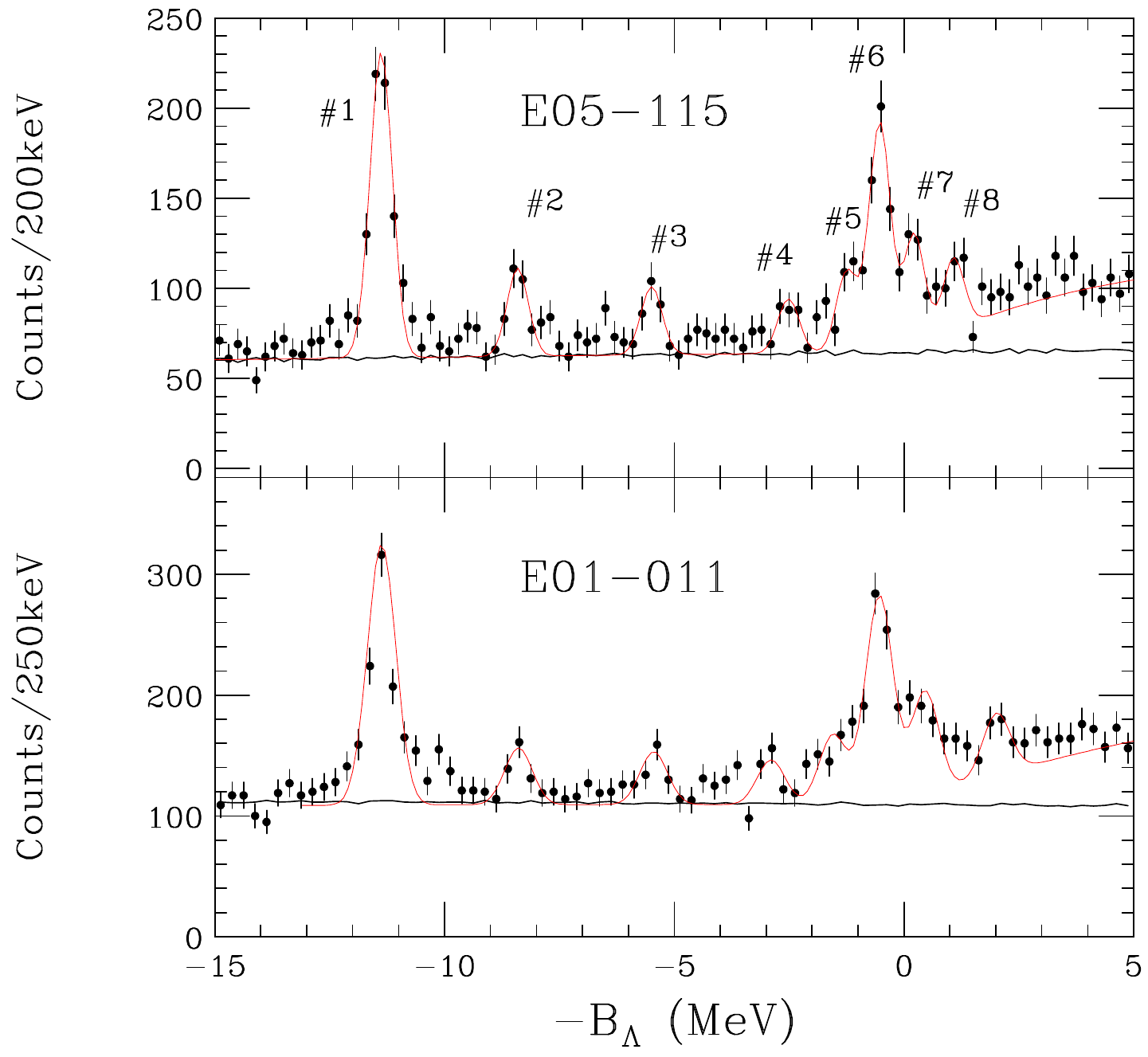}
\caption{\label{fig:figure10} (Color online) Spectroscopy of \lamb{12}{B} 
from the E05-115 and E01-011 experiments. The area below the black line 
is the accidental background.}
\end{figure}

\section{Spectroscopy of $^{\bm {12}}_{\bm{~\Lambda}}$B and Energy Resolution}

 The \lamb{12}{B} spectra from the JLab Hall C E05-115 and E01-011 experiments 
are shown in Fig.~\ref{fig:figure10}. The accidental background shape was 
obtained from the analysis using randomly mixed events from eight accidental 
coincidence peaks in order to reduce the statistical fluctuation.  The two 
experiments have different kinematics acceptances, mainly due to the two 
different electron spectrometers (Enge and HES). The quasi-free distribution 
is fit by a 3rd order polynomial.
Note that the first break up ($\lamb{12}{B} \rightarrow \lamb{11}{Be} + p$) 
is at $-B_{\Lambda} = \unsim +0.9$~MeV, which is just above the threshold.
All possible states below this threshold have a $\Gamma_{\text{EM}}$ decay width 
which is much smaller than the experimental resolution.  Therefore, all 
structures are expected to have the same width with the exception of 
neighboring doublets which lie within the experimental energy resolution. 

 Eight peaks in each spectrum can be recognized as having a statistical 
significance larger than $4\sigma$.  These are fit by a Gaussian function. 
With the exception of the ground state peak which is obviously broader, the 
least $\chi^2/\text{NDF}$ (or the best confidence level (C.L. $\unsim 90\%$)) 
is obtained by assuming that all of the other seven peaks have the same width.  
The widths of the peaks in the E05-115 experiment are 
$\sigma = 231\pm 30~\text{keV}$, while those in the E01-011 are 
$\sigma = 300\pm 50~\text{keV}$.  Therefore, the energy resolution is 
confirmed to be $\unsim 540$ keV and $\unsim 710$ keV FWHM for E05-115 and 
E01-011, respectively.

 Using a single Gaussian fit, the ground state peak is found located at 
$B_{\Lambda} = 11.380\pm 0.020~\text{MeV}$ with a width of 
$\sigma = 271\pm 21~\text{keV}$ in the E05-115 spectrum and at
$B_{\Lambda} = 11.379\pm 0.026~\text{MeV}$ with a width of 
$\sigma = 339\pm 33~\text{keV}$ in the E01-011 spectrum. Though a clear 
separation of the ground-state doublets is difficult without any constraints, 
a double-Gaussian fit study was carried out with a fixed energy resolution as 
described above and a peak amplitude constraint based on a theoretical 
prediction. The cross-section ratio ($2^-_1$/$1^-_1$) for producing the 
doublet states has been estimated to be $\unsim 3.6$ at small 
angles~\cite{Iodice07,Motoba10}. The estimated ratio depends on the interaction models 
used for the calculation, however, the population of the $2^-_1$ state
is always expected to be 3-4 times larger than that of the $1^-_1$ state in 
the HKS kinematics. Therefore, we constrained our double-Gaussian fit with a
peak amplitude ratio of 3.5 with a single free parameter for the peak separation.
The fit gave the peak separations of the doublet as $181 \pm 25~\text{keV}$ 
for the E05-115 spectrum and $176 \pm 31~\text{keV}$ for the E01-011 spectrum.  
The statistical uncertainty of the positions of these peaks is $\pm 17~\text{keV}$ 
and $\pm 24~\text{keV}$ for the E05-115 and E01-011 spectra respectively. 
These are consistent and their weighted average gives a separation of 
$179 \pm 19~\text{keV}$.

 Table~\ref{tab:table3} lists the positions of the eight fitted peaks in terms 
of binding energy $B_{\Lambda}$.  The average values are also given. Although 
the E01-011 spectrum has lower resolution and higher statistical uncertainty 
in comparison to that from E05-115, the results are consistent when accounting 
for the statistical and systematic uncertainties.  

 The photo-production cross sections are obtained~\cite{Nakamura13} using
the virtual photon flux ($\Gamma$)
\begin{equation}
  \label{eqn:eqn2}
  \frac{d\sigma}{d\Omega_K} = \frac{1}{\Gamma} \frac{d \sigma}{dE_{e'} 
d\Omega_{e'} d\Omega_{K}}\ ,
\end{equation}
where the virtual photon flux is integrated over the momentum and angular 
acceptances of the electron spectrometer (HES for E05-115 and Enge for
E01-011). The cross sections of the corresponding peaks from the two 
independently obtained spectra are listed separately without average. 
Notice that because the E05-115 experiment had a larger mean 
$\theta_{\gamma K}$ angle (Table~\ref{tab:table1}), its measured cross 
sections are expected to be lower. The systematic uncertainty of the 
experimentally obtained cross sections is about $\pm 12\%$. 

\begin{table*}[htp]
\begin{threeparttable}
\caption{\label{tab:table3} Binding energies and cross sections
  of the fitted peaks.  The uncertainty listed in
  table is statistical. The systematic uncertainty for $B_{\Lambda}$
  is $\pm0.11~\text{MeV}$ and $\pm 0.16~\text{MeV}$ for E05-115
  and E01-011, respectively.  This systematic uncertainty mainly
  causes a shift for $B_{\Lambda}$ of all the peaks.}
\begin{ruledtabular}
\begin{tabular}{crrrcc}
\multirow{2}{*}{Peak} & $B_{\Lambda}$ (MeV) & $B_{\Lambda}$ (MeV) &
$B_{\Lambda}$ (MeV) & Cross Section (nb/sr) & Cross Section (nb/sr) \\
 & (E05-115) & (E01-011) & Average & (E05-115) & (E01-011)\\
\hline
\#1-1\tnote{$\dagger$}&$11.529\pm 0.025$ & $11.517\pm 0.031$ & $11.524\pm 0.019$ &
\multirow{2}{*}{$83.0\pm 3.0$} & \multirow{2}{*}{$101.0\pm 4.2$} \\
\#1-2\tnote{$\dagger$}&$11.348\pm 0.025$ & $11.341\pm 0.031$&$11.345\pm 0.019$ & &\\
\#2 & $8.425\pm 0.047$ & $8.390\pm 0.075$ & $8.415\pm 0.040$ &
$19.1\pm 3.7$ & $\phantom{11}33.5\pm 11.3$ \\
\#3 & $5.488\pm 0.052$ & $5.440\pm 0.085$ & $5.475\pm 0.044$ & 
$18.0\pm 4.6$ & $\phantom{1}26.0\pm 8.8$ \\
\#4 & $2.499\pm 0.075$ & $2.882\pm 0.085$ & $2.667\pm 0.056$ &
$16.2\pm 5.1$ & $\phantom{1}20.5\pm 7.3$ \\
\#5 & $1.220\pm 0.056$ & $1.470\pm 0.091$&$1.289\pm 0.048$ &
$28.7\pm 7.2$ & $\phantom{1}31.5\pm 7.4$ \\
\#6 & $0.524\pm 0.024$ & $0.548\pm 0.035$ & $0.532\pm 0.020$ & 
$\phantom{1}75.7\pm 10.8$ & $\phantom{11}87.7\pm 15.4$ \\
\#7 & $-0.223\pm 0.039$ & $-0.318\pm 0.085$ & $-0.240\pm 0.035$ &
$39.0\pm 7.4$ & $\phantom{11}46.3\pm 10.3$ \\
\#8 & $-1.047\pm 0.078$ & $-0.849\pm 0.101$ & $-0.973\pm 0.062$ &
$27.8\pm 7.9$ & $\phantom{1}28.5\pm 7.4$ \\
\end{tabular}
\end{ruledtabular}
\begin{tablenotes}
\item [$\dagger$] Separation of the \#1-1 and \#1-2 states was performed
with a fitting constraint on the peak area ratio of 1:3.5.
\end{tablenotes}
\end{threeparttable}
\end{table*}

\subsection{States with a $\Lambda$ in the $s$-shell coupled to the 
low-lying {\Beleven} core states}
\label{sec:4A}

\begin{figure}[th]
\includegraphics[width=8.5cm]{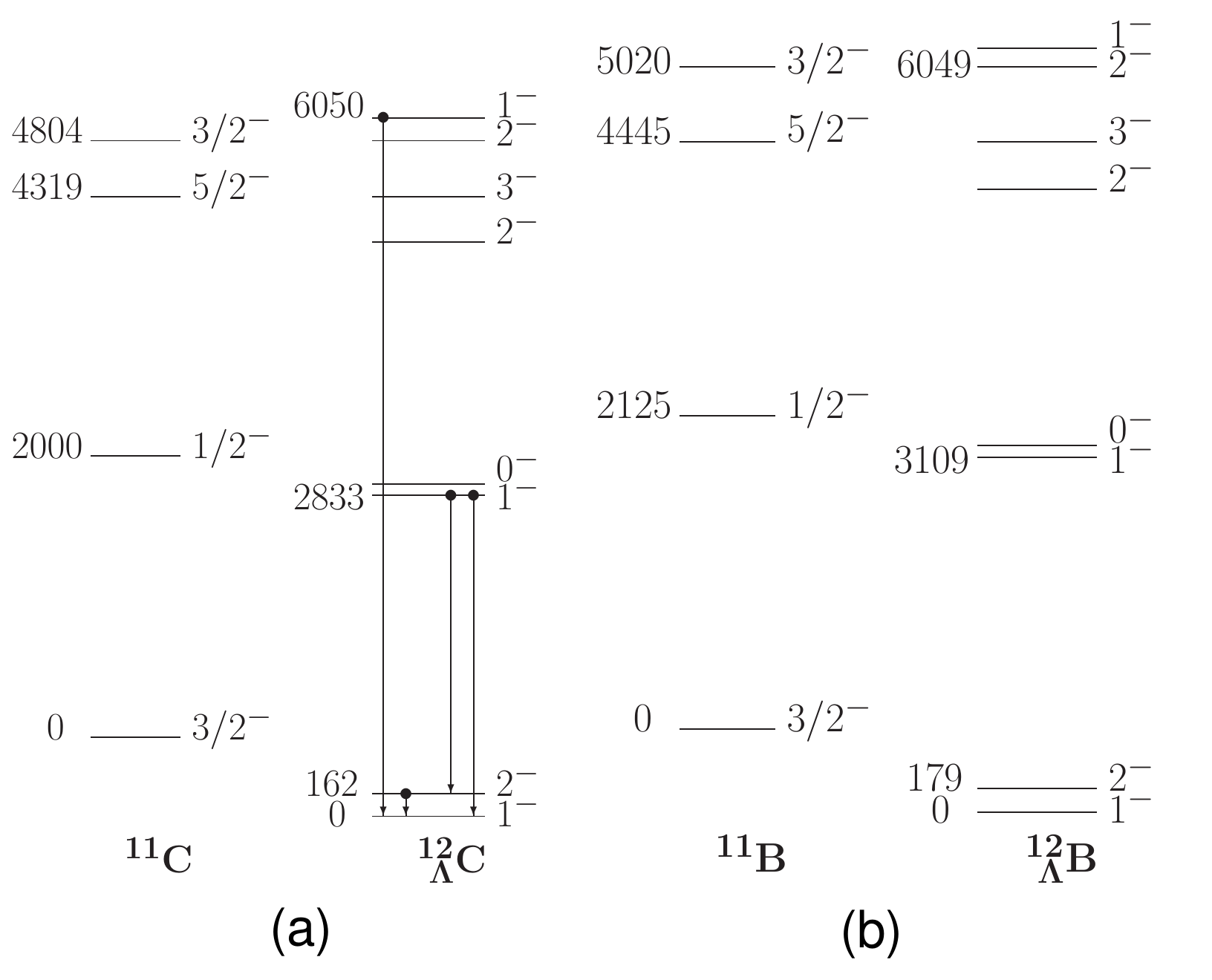}
\caption{\label{fig:figure11} The level structures of \lamb{12}{C} (a) and 
\lamb{12}{B} (b) for the first four doublets (energies in keV). The levels of 
the core nuclei, {\Celeven} (a) and {\Beleven} (b), are also shown. The four 
observed $\gamma$ ray transitions for \lamb{12}{C}~\cite{Hosomi13} are shown 
together with the three deduced excitation energies. For comparison, the 
energies of the peaks \#1-2, \#2, and \#3 are shown for \lamb{12}{B} 
(Tables~\ref{tab:table3}).}
\end{figure}

 The peaks from \#1 (containing 1-1 and 1-2) to \#4 are all considered to have 
a $\Lambda$ in an $s$-shell which is coupled to the {\Beleven} core.  Peak \#4 
will be discussed later and peaks \#1 to \#3 are considered to have a 
negative parity core structure.  Peak \#1-1 and \#1-2 are the ground state 
doublet states $1^-_1$ and $2^-_1$ with a $\Lambda_s$ coupled to the $3/2^-$ 
{\Beleven} ground state.  Peak \#2 is considered to be the lower member of 
the second doublet ($1^-_2$ and $0^-_1$) with the $1/2^-$ core, while
peak \#3 should be the lower member of the fourth doublet ($2^-_3$ and $1^-_3$) 
with the second $3/2^-$ {\Beleven} core.  The $0^-_1$ and $1^-_3$, as well as 
the third doublet ($2^-_2$ and $3^-_1$), are all predicted to have small cross 
sections ($<$ few nb/sr) and thus are difficult to observe without sufficient
statistics and a better signal/background ratio.  Using the averaged 
$B_\Lambda$ values from the two experiments, the assumed level structures 
are illustrated in Fig.~\ref{fig:figure11}(b) in terms of the excitation 
energy spectrum with respect to the ($1^-_1$) ground state.  The systematic 
uncertainty for these extracted excitation energies is $\unsim\pm 0.07$ MeV.

 This observed excitation level spectrum of \lamb{12}{B} with a $\Lambda$ in 
$s$-shell can be compared to that of the mirror hypernucleus \lamb{12}{C}  
(shown in Fig.~\ref{fig:figure11}(a)) which was constructed by four 
precisely measured $\gamma$ transitions~\cite{Hosomi13}.  Other than small 
excitation energy differences, {\Beleven} and {\Celeven} have the same level 
structure. To a first approximation, the excitation energies of states in the 
excited doublets of \lamb{12}{B} can be obtained by adding the difference 
between the excitation energies of the core states of {\Beleven} and {\Celeven}.

 The ground state doublet separation for \lamb{12}{C} was measured to be 
0.162 MeV, while the separation of the same doublet for \lamb{12}{B} measured 
by the HKS experiments is 0.179 MeV.  If simply taking into account the 
2.125 - 2.000 = 0.125 MeV energy difference between the first excited 
$1/2^-$ states of {\Beleven} and {\Celeven}, the $1^-_2$ state for 
\lamb{12}{B} in the second doublet can be predicted to have a
2.958 MeV excitation which is close to the measured value of 3.109 MeV.  When 
comparing the fourth doublet, the $\gamma$ transition measured by the KEK 
experiment was assigned to lie between the states of $1^-_3$ and $1^-_1$.  
However, in the $(e,e'K^+)$ reaction the lower member of the doublet, the 
$2^-_3$ state, is favored. Thus the two experiments should have measured 
different states in the same doublet. By simply adding the energy difference 
of $5.020\!-\!4.804\!=\!0.216$ MeV, the upper member (i.e. the $1^-_3$ state) 
is suggested to have excitation energy of 6.266 MeV. Thus, the simple estimate 
gives the separation of the two states in the fourth doublet on the order of 
$\unsim 0.22 \pm 0.09 {\rm stat.} \pm 0.07 {\rm sys.}$ MeV. The recent 
theoretical calculations are 0.107 MeV as presented in Ref.~\cite{Iodice07} and 
0.122 MeV from a G-matrix calculation~\cite{Motoba10}.

\subsection{States from $\Lambda$ in $p$-shell}

 Above the four $s_{\Lambda}$ states (including the $1_1^-$ and $2_1^-$ ground 
state doublet) there are five peaks (\#4 to \#8) with the resolutions (540 keV 
and 710 keV FWHM).  Their averaged excitation energies are 8.86, 10.24, 10.99, 
11.76, and 12.50 MeV with a systematic uncertainty of $\pm 0.07~\text{MeV}$. 

Note that the width of the \#4 peak in the published JLab Hall A spectrum 
(see Table~\ref{tab:table1} in Ref.~\cite{Iodice07}) was fit to be $0.93\pm 0.46~\text{MeV}$ 
FWHM, wider than the reported resolution of 670 keV FWHM.  This excitation was 
reported as $E_x = 9.54\pm 0.16~\text{MeV}$, and there was still unexplained
additional strength.  This result appears to be the average ($E_x \approx
9.55~\text{MeV}$) of peak \#4 and \#5 from the HKS result. In the HKS spectrum, 
peak \#5 peak would then be the first ($2_1^+$ and $1_1^+$) pair of $p_\Lambda$ 
states with a $\Lambda$ in $p$-shell coupled to the $3/2^-$ ground state 
{\Beleven} core.  The energy separation of this doublet (peak \#5) is predicted 
to be very small ($\unsim 40$ keV).

 Peak \#6 is the dominant peak among the $p_\Lambda$ states and is consistent with 
the observation made by the JLab Hall A experiment. This peak can be interpreted 
as the second ($2_2^+$ and $3_1^+$) pair of $p_\Lambda$ states (see theory 
prediction in Table~\ref{tab:table4}). Peak \#8 is located near the first 
breakup threshold and could be the third ($2_3^+$ and $1_2^+$) pair of 
$p_\Lambda$ states.  The width uncertainty for the fitting is about 
$\pm 50~\text{keV}$, which means separation of these two doublet states is 
small, possibly $\le 100~\text{keV}$. 
Peak \#7 appears to be an ``extra'' state and is not predicted by $0\hbar\omega$
based calculations using a $p$-shell core. The peak also exists
in the JLab Hall A spectrum behind the dominant $p$-shell peak.  However, in 
the Hall A analysis the strength was simply fit by one peak with a width of 
1.58 MeV, more than twice the reported resolution (670 keV FWHM).  Thus, the 
peaks \#5, \#6 and \#8 are considered to be three pairs of $p_{\Lambda}$ states.
The possible state configurations and excitation energies are listed in
Table~\ref{tab:table4} and compared to the theoretical calculation used in
Ref.~\cite{Iodice07}. The configuration with $p_{\Lambda}$ infers a strong mixing
of $p_{3/2\Lambda}$ and  $p_{1/2\Lambda}$. The theoretical calculation for the 
$p_{\Lambda}$ states using the G-matrix approach with a ``realistic'' $YN$
interaction~\cite{Motoba10} predicts a different excitation spectroscopy for
the same $p_{\Lambda}$ configuration. A detailed comparison to the present results
for the excitation energies and relative crosss sections may help to improve
the $YN$ interaction model.

\begin{table*}[t]
\caption{\label{tab:table4} Comparison of the measured excitation energies of 
peaks \#5, \#6 and \#8 believed to be based on $p$-shell states of the {\Beleven}
core to the theory calculation of Ref.~\cite{Iodice07}. The main structures of
the theoretical states are given in the second column, with the designation 
$p_\Lambda$ meaning that there is strong mixing of $p_{3/2\Lambda}$ and 
$p_{1/2\Lambda}$ configurations. The $1^+_2$ and $0^+$ states are not expected 
to be seen due to small cross sections and are omitted. The systematic 
uncertainty of the measured excitation energies is about $\pm0.07~\text{MeV}$.}
\begin{ruledtabular}
\begin{tabular}{clccc}
Peak & Structure & $J^{\pi}_n$ & Measured $E_x$ (MeV) &
 Calculated $E_x$~\cite{Iodice07} (MeV) \\
\hline
\multirow{2}{*}{\#5} & $\Beleven(3/2^-; \text{g.s.})\otimes p_{3/2\Lambda}$ &
$2_1^+$ & \multirow{2}{*}{$10.24\pm 0.05$} & 10.29\\
 & $\Beleven(3/2^-; \text{g.s.})\otimes p_{\Lambda}$ & $1_1^+$ & & 10.34\\
\multirow{2}{*}{\#6} & $\Beleven(3/2^-; \text{g.s.})\otimes p_{1/2\Lambda}$ &
$2_2^+$ & \multirow{2}{*}{$10.99\pm 0.03$} & 10.93\\
 & $\Beleven(3/2^-; \text{g.s.})\otimes p_{3/2\Lambda}$ & $3_1^+$ & & 11.01\\
\multirow{2}{*}{\#8} & $\Beleven(1/2^-; 2.125)\otimes p_{3/2\Lambda}$ &
$2_3^+$ & \multirow{2}{*}{$12.50\pm 0.07$} & 12.80\\
 & $\Beleven(1/2^-; 2.125)\otimes p_{\Lambda}$ & $1_3^+$ & & 12.91\\
\end{tabular}
\end{ruledtabular}
\end{table*}

\subsection{$SD$-shell core states with a $\Lambda$ in the $s$-shell}

 The extra peaks \#4 ($E_x = 8.86\pm 0.06~\text{MeV}$) and \#7 
($E_x = 11.75\pm 0.04~\text{MeV}$) may not actually be so surprising.  
Comparisons of early and recent spectroscopic investigations on 
\lamb{12}{C}~\cite{Hashimoto06,Hotchi01,Agnello05} and \lamb{12}{B}~\cite{Iodice07} 
with $0\hbar\omega$ shell-model calculations commonly result in leftover strength 
around the $p_\Lambda$ peaks.  In the HKS experiments, these are fit using the 
same width as the other peaks.  As discussed also in Ref.~\cite{Iodice07}, these 
two peaks may be due to states with a configuration of $s_{\Lambda}$ coupled to 
the $3/2^+$ and $5/2^+$ $sd$-shell {\Beleven} core states.  Their excitation 
energies would happen to be near the strong $p_\Lambda$ states and they would 
get their strength from mixing with these states.  Theoretical investigations 
using a full $1\hbar \omega$ basis are indeed needed. 

\section{Conclusions}

 The unique CEBAF beam has enabled high precision spectroscopic investigation 
of $\Lambda$ hypernuclei which are a laboratory for studying $\Lambda N$ 
interactions. The independent Hall C HKS experiments, E05-115 and E01-011, 
provide consistent results.  They obtained excellent energy resolution which 
is essential for obtaining the detailed level structures presented here. This
paper demonstrates how to calibrate a system of two spectrometers in which the
angle and momentum reconstruction matrices are coupled by using the
calibration data obtained from {\eepK} reaction. In addition, the analysis 
of the \lamb{12}{B} hypernuclear spectra using the confirmed energy resolution
($\unsim 540$ keV for E05-115 and $\unsim 710$ keV for E01-011) has revealed 
new states and determined the ground state mass. The experiments have also 
confirmed the existing level and spin structure of this hypernucleus. The 
observed states provide a challenge for theoretical calculations. Future 
technical improvements will seek to reduce the high accidental background.

\begin{acknowledgments}
We acknowledge continuous support and encouragement from the staff of
the Jefferson Lab physics and accelerator divisions. The hypernuclear
programs at JLab Hall-C are supported by Japan-MEXT Grant in-aid for
Scientific Research (16GS0201, 15684005,12002001, 08239102, 09304028,
09554007, 11440070, 15204014), Japan-US collaborative research
program, Core-to-core program (21002) and strategic young researcher
overseas visits program for accelerating brain circulation (R2201) by
JSPS, US-DOE contracts (DE-AC05-84ER40150, DE-AC05-06OR23177, DE-FG02-99ER41065,
DE-FG02-97ER41047, DE-AC02-06CH11357, DE-FG02-00ER41110 and DE-AC02-98CH10886) and 
US-NSF contracts (013815 and 0758095).
\end{acknowledgments}

\end{document}